\documentclass[prd,showpacs,nofootinbib,preprint,10 pt]{revtex4}
\usepackage{amsmath,graphicx,color,xcolor,epsfig}
\usepackage{epstopdf}
\usepackage{hyperref}

\def\Q{{\mathcal{Q}}}
\def\P{{\mathcal{P}}}

\begin{document}
\begin{titlepage}

\begin{flushright}
DESY 16-102\\
June 2016
\end{flushright}

\begin{center}
\setlength {\baselineskip}{0.3in} 
{\bf\Large\boldmath Heavy quark symmetry and weak decays of the $b$-baryons in pentaquarks with a $c\bar{c}$ component.}\\[5mm]
\setlength {\baselineskip}{0.2in}
{\large  Ahmed Ali$^1$, Ishtiaq Ahmed$^{2}$, M. Jamil Aslam$^{3}$ and~ Abdur Rehman$^2$}\\[5mm]

$^1$~{\it Deutsches Elektronen-Synchrotron DESY, D-22607 Hamburg, Germany.}\\[5mm]

$^2$~{\it National Centre for Physics, Quaid-i-Azam University Campus,\\
             Islamabad 45320, Pakistan.}\\[5mm]
$^3$~{\it Physics Department, Quaid-i-Azam University,\\ Islamabad 45320, Pakistan.}\\[3cm] 
\end{center}

{\bf Abstract}\\[5mm] 
\setlength{\baselineskip}{0.2in} 
The discovery of the baryonic states $P_c^+(4380)$ and $P_c^+(4450)$  by the
 LHCb collaboration in the process $p p \to b\bar{b} \to \Lambda_b^0 X$, followed by the decay
 $\Lambda_b^0 \to J/\psi\; p\; K^-$ has evoked a lot of
 theoretical interest. These states 
have the minimal quark content $c \bar{c} u u d$, as suggested by their discovery mode
$J/\psi \; p$, and the preferred $J^P$ assignments are $\frac{5}{2}^+$ for the
$P_c^+(4450)$ and $\frac{3}{2}^-$ for the $P_c^+(4380)$.  In the compact pentaquark hypothesis, 
in which they are interpreted as hidden charm 
diquark-diquark-antiquark baryons, the assigned spin and angular momentum quantum numbers are
$P_c^+(4380)= \{\bar{c} [cu]_{s=1} [ud]_{s=1}; L_{\mathcal{P}}=0,
J^{\rm P}=\frac{3}{2}^- \}$ and $P_c^+(4450)= \{\bar{c} [cu]_{s=1} [ud]_{s=0}; L_{\mathcal{P}}=1,
J^{\rm P}=\frac{5}{2}^+ \}$. The subscripts denote the spin of the diquarks and $L_{\mathcal{P}}=0,1$ are the orbital angular momentum 
quantum  numbers of the pentaquarks. We point out that in the heavy quark limit, the spin of the light diquark in 
heavy baryons  becomes a good quantum number, which has consequences for the
decay $\Lambda_b^0 \to J/\psi \; p\; K^-$. With the quantum numbers
assigned above for the two pentaquarks, this would allow
only the higher mass pentaquark state $P_c^+(4450)$  having $[ud]_{s=0}$ to be produced in $\Lambda_b^0$ decays, 
whereas the lower mass state  $P_c^+(4380)$ having $[ud]_{s=1}$ is disfavored, requiring a different
interpretation. Pentaquark spectrum is  rich enough to accommodate a $J^P=\frac{3}{2}^-$ state, which has the correct light diquark spin $\{\bar{c} [cu]_{s=1} [ud]_{s=0}; L_{\mathcal{P}}=0, J^{\rm P}=\frac{3}{2}^- \}$ to be produced in $\Lambda_b^0$ decays.
Assuming that the mass difference between the charmed pentaquarks which differ in the orbital angular momentum 
$L$ by one unit is similar to the corresponding mass difference in the charmed baryons,  $m[\Lambda_c^+(2625); J^P=\frac{3}{2}^-] - m[\Lambda_c^+(2286); J^P=\frac{1}{2}^+] \simeq 341$ MeV,  we estimate the mass of the lower pentaquark $J^P=3/2^-$ state to be  about 4110  MeV and  suggest to reanalyze the LHCb data to search for
 this third state. 
 Extending these considerations to the pentaquark states having a $c \bar{c}$ pair and  three light quarks ($u$, $d$,
 $s$) in their Fock space,  we present  the spectroscopy of the $S$- and $P$-wave states in
 an effective Hamiltonian approach.  Some of these pentaquarks can be produced in weak decays of  the $b$-baryons.
 Combining heavy quark symmetry and the $SU(3)_F$ symmetry results in
 strikingly simple relations among the decay amplitudes which are presented here.

\end{titlepage}

\section{Introduction}

The discovery of the charmonium-like resonance $X(3872)$ in $B$-meson decays
$B \to X(3872)\; K$, followed by the decay $X(3872) \to J/\psi \pi^+\pi^-$, reported by the Belle 
collaboration in 2003~\cite{Belle-C}, subsequently
confirmed by the D0~\cite{D0}, CDF~\cite{ CDF II} and Babar Collaborations~\cite{Babar}, has proved to be
the harbinger of a new quarkonium-like spectroscopy. Since then, well over two dozen such hidden
$c\bar{c}$ states, both neutral and charged, have been reported. Very recently, observation of four
structures in the $J/\psi \phi$ mass spectrum in the decays $B^+ \to J/\psi\; \phi \; K^+$ have been reported
by LHCb, yielding two $J^P=1^+$ states, $X(4140)$ and $X(4274)$, and two $J^P=0^+$ states
$X(4500)$ and $X(4700)$ \cite{Aaij:2016iza}.  So far, three states 
$Y_b(10890)$~\cite{Abe:2007tk}, $Z_b^\pm (10610)$ and $Z_b^\pm (10650)$~\cite{Belle:2011aa} have also been 
discovered having a $b\bar{b}$ pair in their valence compositions.
% (For an experimental review,  see ~\cite{Olsen:2015zcy}).
 All these hadrons
are distinct by the presence of a $c\bar{c}$ (or a $b\bar{b}$) quark pair in
addition to light degrees of freedom (a light $q\bar{q}$ pair or gluons) in their Fock space.
Collectively called the $X, Y,  Z$ states, they have prompted a lot of
theoretical interest in their interpretations~\cite{Maiani:2004vq} - \cite{Chen}.
In 2015, LHCb  reported the first observation of two hidden charm pentaquark states $P_{c}^+(4380)$ and $P_c^+(4450)$ in the decay $\Lambda_b^0 \to J/\psi\; p\; K^-$ ~\cite{Aaij:2015tga}, having the masses
$4380 \pm 8 \pm 29$ MeV and $4449.8 \pm 1.7 \pm 2.5$ MeV, and widths $205 \pm 18 \pm 86$ MeV
and $39 \pm 5 \pm 19$ MeV, with the
preferred spin-parity assignments $J^{P}=\frac{3}{2}^-$ and $J^P=\frac{5}{2}^+$, respectively.
 These states have the quark composition $c\bar{c} uud$, and like their tetraquark
counterparts $X, Y,  Z$, they  lie close in mass to several (charm meson-baryon) thresholds. This has led to a number of theoretical proposals for their interpretation, which include  rescattering-induced kinematical effects \cite{rescattering}, open charm-baryon and  charm-meson bound states \cite{meson-boundstates}, and baryocharmonia \cite{Baryocharmonia}. They have also been interpreted as compact pentaquark hadrons with the internal structure organized as diquark-diquark-anti-charm quark~\cite{Maiani:2015vwa,compact-Pentaquark}
or as diquark-triquark~\cite{Lebed:2015tna,Zhu:2015bba}.  

In this work we follow the compact pentaquark interpretation. The 
basic idea is that highly correlated diquarks play a key role in the physics of
multiquark states~\cite{Lipkin:1987sk,Jaffe:2003sg,Maiani:2004vq}. 
 Since quarks transform as a triplet $\tt 3$ of color $SU(3)$, the diquarks resulting from the
direct product $\tt 3 \otimes 3=\bar{3} \oplus 6$, are thus either a color anti-triplet $\tt \bar{3}$ or a
color sextet $\tt 6$. Of these only the  color $\tt \bar{3}$ configuration is kept,
 as suggested by perturbative arguments. Both spin-1 and spin-0 diquarks are, however, allowed.
In the case of a diquark $[qq^\prime]$ consisting of two light quarks, the spin-0 diquarks are believed to be more
tightly bound than the spin-1, and this hyperfine splitting has implications for the spectroscopy.
For the heavy-light diquarks, such as $[cq]$ or $[bq]$, this splitting is suppressed by $1/m_c$
for a $[cq]$ or by $1/m_b$ for a $[bq]$ diquark, and hence both spin-configurations are treated at par. 
For the pentaquarks, the mass spectrum  will depend upon how the five quarks, i.e., the 4 quarks and an antiquark, are dynamically  structured. A  diquark-triquark picture, in which the two observed pentaquarks consist of 
a rapidly separating pair
of a color-${\tt \bar{3}}$ $[cu]$ diquark and a color-${\tt 3}$ triquark $\bar{\theta}=\bar{c}[ud]$, has been
presented in~\cite{Lebed:2015tna}. A ``Cornell''-type non-relativistic linear-plus-coulomb potential \cite{Eichten:1978tg}  is  used to determine the diquark-triquark separation $R$ and the ensuing phenomenology is worked out. 

\begin{figure}
\centerline{\includegraphics[scale=0.25]{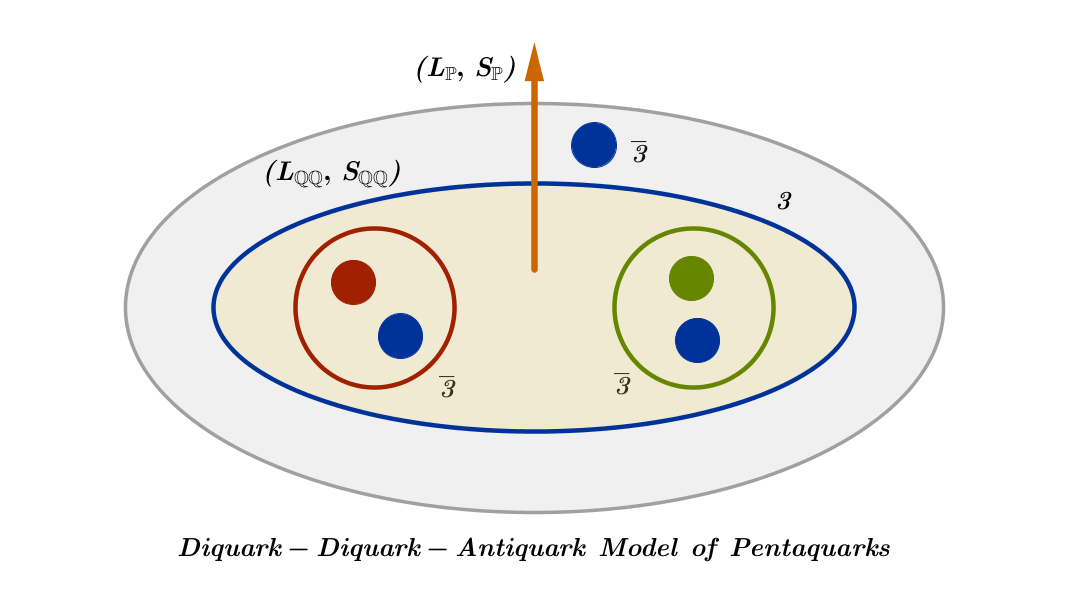}}
\vspace*{-4mm}
\caption{$SU(3)$-color quantum numbers of the diquarks, tetraquark and antiquark are indicated, together
with the orbital and spin quantum numbers of the tetraquark and pentaquark~\cite{Ali:2016gli}}.
\label{aaar:fig1}
\end{figure}

We prefer to keep the basic building blocks of the pentaquarks 
to be quarks and diquarks, and  follow  here the template in which the
5q baryons, such as the two $P_c$ states, are assumed to be four quarks, consisting of two highly
correlated diquark pairs, and an antiquark. For the  present discussion, it is an anti-charm quark $\bar{c}$ which
is correlated with the two
diquarks  $[cq]$ and $[q^{\prime}q^{\prime \prime}]$, where $q,q^{\prime},q^{\prime \prime}$ can be $u$ or $d$.
The tetraquark formed by the diquark-diquark $([cq]_{\tt \bar{3}}[q^{\prime}q^{\prime \prime}]_{\tt \bar{3}})$ is a color-triplet
object, following from ${\tt \bar{3} \times \bar{3} = \bar{6} + 3}$, with orbital and spin quantum numbers,
 denoted by $L_{\Q\Q}$ and $S_{\Q\Q}$, 
which combines with the color-anti-triplet ${\tt \bar{3}}$ of the
 $\bar{c}$  to form an overall color-singlet pentaquark,
with the corresponding quantum numbers $L_{\P}$ and $S_{\P}$. This is 
shown schematically in Fig.~\ref{aaar:fig1}.
%\vspace*{-5mm}

An effective Hamiltonian based on this picture is constructed, extending the underlying tetraquark 
Hamiltonian developed for the $X,Y, Z$ states~\cite{Maiani:2004vq}. We
explain how the various input parameters  in this Hamiltonian are determined. Subsequently, we work out the mass spectrum of the low-lying $S$- and $P$-wave 
pentaquark states, with a $c\bar{c}$ and three light quarks ($u$, $d$, $s$) in their Fock space. 

The pentaquark states reported by the LHCb are produced in  $\Lambda_b^0$ decays, $\Lambda_b^0 \to \P^{+}\;K^{-}$, where $\P$ denotes a generic pentaquark state, a symbol we use subsequently in this work. We take a
closer look at the dynamics of $\Lambda_b$ decays. In particular, we point out that
QCD has a symmetry in the heavy quark limit, i.e., for
 $m_b \gg \Lambda_{\rm QCD}$, $b$-quark becomes a static quark and the light diquark spin becomes a
 good quantum number, constraining the states which can otherwise be produced in $b$-baryon decays. 
 The consequences of heavy quark symmetry are well known, starting from the early uses 
in the decays of the heavy mesons ($B, B^*$ etc.)~\cite{Isgur:1989vq,Isgur:1989ed},
for the heavy meson spectroscopy~\cite{Isgur:1991wq}, 
 and in heavy baryon  decays~\cite{Mannel:1990vg,Falk:1991nq,Manohar:2000dt}.
 The extent to which heavy quark symmetry  holds can be judged from the data on 
 the semileptonic decays $\Lambda_b^0 \to \Lambda_c^+ \ell^-\bar{\nu}_\ell$, 
 which is a $j^P=0^+ \to j^P=0^+$ transition,
 for which a branching ratio
${\cal B}(\Lambda_b^0 \to \Lambda_c^+ \ell^-\bar{\nu}_\ell)= (6.2 ^{+1.4}_{-1.2})\%$
is listed in the PDG~\cite{Agashe:2014kda}. The corresponding decay
 $ \Lambda_b^0 \to \Sigma_c^+ \ell^-\bar{\nu}_\ell $, involving a $j^P=0^+ \to j^P=1^+$ transition,
 is non-existent. The  decays
$\Lambda_b^0 \to \Sigma(2455)^0 \pi^+ \ell^-\bar{\nu}_\ell$ and $\Lambda_b^0 \to \Sigma(2455)^{++} \pi^- \ell^-\bar{\nu}_\ell$, facilitating an $0^+ \to 1^+$ transition, are highly suppressed,
$ (1/2 \Gamma (\Lambda_b^0 \to \Sigma(2455)^0 \pi^+ \ell^-\bar{\nu}_\ell ) + 1/2 \Gamma(\Lambda_b^0 \to \Sigma(2455)^{++} \pi^- \ell^-\bar{\nu}_\ell)/\Gamma(\Lambda_b^0 \to \Lambda_c^+ \ell^-\bar{\nu}_\ell  ) 
=0.054 \pm 0.022 ^{+0.021}_{-0.018} $~\cite{Agashe:2014kda}.
 For the non-leptonic decays, one finds, for example,  
${\cal B}(\Lambda_b^0 \to \Sigma_c^0(2455) \pi^+\pi^-)/ {\cal B} (\Lambda_b^0 \to \Lambda_c^+ \pi^-)  \simeq 0.1$, indicating an order of magnitude suppression of  the $j^P=0^+ \to j^P=1^+$ transition.
  Whether the heavy quark symmetry holds in $b$-baryon decays to pentaquarks  is, of course, a
dynamical question and we currently lack data to test it, but it is worthwhile to work out its implications for
the interpretation of the LHCb data and the pentaquark phenomenology, in general.

  In the pioneering work by Maiani~{\it et al.}~\cite{Maiani:2015vwa} on the pentaquark interpretation
  of the  LHCb data on $\Lambda_b^0 \to J/\psi\; p\; K^-$ decay, heavy quark symmetry is not invoked.  The assigned internal quantum numbers are:
  $P_c^+(4450)= \{\bar{c} [cu]_{s=1} [ud]_{s=0}; L_{\mathcal{P}}=1,J^{\rm P}=\frac{5}{2}^+ \}$ 
  and  $P_c^+(4380)= \{\bar{c} [cu]_{s=1} [ud]_{s=1}; L_{\mathcal{P}}=0, J^{\rm P}=\frac{3}{2}^- \}$. 
  Taking into account the mass differences due to the orbital angular momentum and the light diquark spins,
  the observed mass difference between the two $P_c^+$ states of about 70 MeV is reproduced.
  The crucial assumption is that the two diagrams for the decay
 $\Lambda_b^0 \to J/\psi\; p \;K^-$ in Fig.~1 in~\cite{Maiani:2015vwa}, in which the $ud$-spin in
 $\Lambda_b^0$ goes over to the $[ud]$-diquark spin in the pentaquark, Fig.~1(A), and the one in which the 
 $ud$-spin is shared among the final state pentaquark and a meson, generating a light diquark $[ud]$ having 
  spin-0 and spin-1, Fig.~1(B), are treated at par. This, as we discussed in the previous paragraph, is at variance
  with the data on $b \to c$ baryonic decays, and also with the heavy quark symmetry.

 We argue here that the $b$-baryon decays to pentaquarks
 having a  $c\bar{c}$ component are also subject to the selection rules following from the heavy quark
 symmetry. In particular, this
 implies a dynamical suppression of  Fig.~1(B) in~\cite{Maiani:2015vwa}.  With this additional symmetry as a diagnostic tool, we analyze the two observed pentaquark states in
 $\Lambda_b^0$ decays and find that only the higher mass sate $P_c^+(4450)= \{\bar{c} [cu]_{s=1} [ud]_{s=0}; L_{\mathcal{P}}=1,J^{\rm P}=\frac{5}{2}^+ \}$ is allowed to be produced in $\Lambda_b^0$ decays, but the lower-mass
 state $P_c^+(4380)= \{\bar{c} [cu]_{s=1} [ud]_{s=1}; L_{\mathcal{P}}=0, J^{\rm P}=\frac{3}{2}^- \}$, in
 which the spin-0 $(ud)$ diquark in $\Lambda_b^0$ is broken,  is disfavored 
 in this limit.  Interestingly, the fractions of the total sample in the decay $\Lambda_b^0 \to J/\psi \; p\;K^-$ due to $P_c^+(4380)$ and $P_c^+(4450)$ are
 reported by the LHCb~\cite{Aaij:2015tga}  as $(8.4 \pm 0.7 \pm 4.2)$\% and  $(4.1 \pm 0.75 \pm 1.1)$\%, respectively.
 Thus, another theoretical interpretation may be required to accommodate the state $P_c^+(4380)$ in the LHCb data.

Spectrum of the multiquark states is, however, rich, as also presented here for the $S$- and $P$-wave pentaquark states having a hidden $c\bar{c}$ pair and three light ($u$, $d$, $s$) quarks.
We find that there indeed is  a lower-mass
  $J^P=\frac{3}{2}^-$ pentaquark state  with the quantum numbers $\{\bar{c} [cu]_{s=1} [ud]_{s=0}; L_{\mathcal{P}}=0, J^{\rm P}=\frac{3}{2}^- \}$
present  in the spectrum, which has the correct light diquark spin  to be produced in the decay
$\Lambda_b^0 \to J/\psi\; p\; K^-$,  compatible with the heavy quark symmetry. 
Assuming that the mass difference in the charmed pentaquarks differing by
the orbital angular momentum $\Delta L =1$  is similar to the corresponding mass difference in the charmed baryons,  $m[\Lambda_c^+(2625); J^P=\frac{3}{2}^-] - m[\Lambda_c^+(2286); J^P=\frac{1}{2}^+] \simeq 341$ MeV,  we predict  the mass of the lower pentaquark state to be about 4110  MeV. Estimates based on the
 parameters of the $c\bar{c}$ tetraquarks in an effective Hamiltonian approach, which we detail in this
paper, yield a nominally larger but compatible value for the mass of this state, around 4130 MeV.  We suggest to search for the lower mass $P_c^+ (J^P = \frac{3}{2}^-)$ state decaying into $J/\psi\; p$  in the  LHCb data
on $\Lambda_b^0 \to J/\psi\; p K^-$. A  renewed fit of the LHCb data by allowing a third  resonance 
is called for.
 
\begin{figure}
\centerline{\includegraphics[scale=0.50]{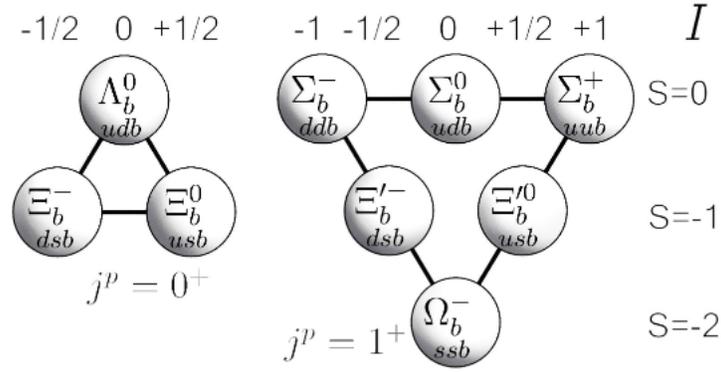}}
\vspace*{-4mm}
\caption{The $SU(3)_F$ flavor multiplets of the ground-state bottom baryons. The
$J^P=\frac{1}{2}^+$ triplet (with the light-quark spin $j^P=0^+$) is shown on the left, and the $J^P=\frac{1}{2}^+$ sextet (with the light-quark spin $j^p=1^+$)
is shown on the right. The isospin $I$ and strangeness  $S$ of each $b$-baryon state are specified (taken from \cite{Ali:2012pn}).}
\label{Fig:b-baryon-multiplet}
\end{figure}

In addition to the $\Lambda_b^0=(udb)$, which is the lightest of the $b$-baryons in which the light quark pair $ub$
has $j^P=0^+$, there are two others in this $SU(3)_F$ triplet with strangeness $S=-1$, 
$\Xi_b^0(5792)=(usb)$, having isospin $I=I_3=1/2$ and  $\Xi_b^-(5794)=(dsb)$, having isospin $I=-I_3=1/2$.
Likewise, there are six $b$-baryons with the light quark pair having $j^P=1^+$,  with $S=0$
 ($\Sigma_b^-=(ddb), \; \Sigma_b^0=(udb), \; \Sigma_b^+=(uub))$, $S=-1$ ($\Xi_b^\prime=(dsb),\;\Xi_b^{\prime 0}=(usb)$),  and one with $S=-2$ ($\Omega_b^-=(ssb) $.) These bottom baryon multiplets are shown in 
Fig. \ref{Fig:b-baryon-multiplet}.

 Weak decays of some of these
$b$-baryons are expected to produce pentaquark states with a hidden $c\bar{c}$ pair
and three light quarks, which form baryons present in  the octet and decuplet representations of
$SU(3)_F$. The observed pentaquarks  $P_{c}^+(4450)$ and the one being proposed here
 $P_c^+(4110)$ belong to the $SU(3)_F$ octet. Examples of the bottom-strange $b$-baryon respecting $\Delta I=0,\, \Delta S=-1$ are~\cite{Maiani:2015vwa}:
 \vspace*{-2mm}
\begin{align}
\Omega_b^- \to K^- (J/\psi\; \Xi^0),\; K^0 (J/\psi\; \Xi^-), \nonumber
\end{align}
which correspond to the formation of the pentaquarks in the $SU(3)_F$ decuplet representation with the spin configuration
$ \P_{10} (\bar{c}\, [cu]_{s=0,1}\, [s s]_{s=1})$.
 Thus, while calculating the decay amplitudes of these transitions is a formidable challenge,
$SU(3)_F$ symmetry  can be used to relate various decays.
This was already pointed out by Maiani \textit{et al.}~\cite{Maiani:2015vwa}, and these relations have been
worked out in great detail subsequently~\cite{Cheng-Chua, Li-He-He}. They provide a very useful guide for the future  pentaquark searches. We point out that imposing the heavy quark symmetry brings clarity in
this analysis, reducing the number of unknown matrix elements and providing a better understanding of why
some topological diagrams are disfavored. In addition,  as already pointed out, heavy quark symmetry
reduces the number of pentaquark states which can be reached in $b$-baryon decays.
 We revisit the decays 
of the $b$-baryons to pentaquarks and pseudoscalar mesons, which were considered in~\cite{Cheng-Chua, Li-He-He} without this symmetry constraint.  

This paper is organized as follows. In section II, we give the effective Hamiltonian used to work out the
pentaquark mass spectrum and specify the values of the various input parameters. Section III  contains our
predictions for the pentaquark masses having the quark flavors $\bar{c}[cq][q^\prime q^{\prime \prime}]$, with
$q, q^\prime, q^{\prime \prime} =u,d,s$, assuming isospin symmetry. In section IV, we discuss decays of the
$b$-baryons into pentaquarks, obeying the heavy quark symmetry, which is followed by the $SU(3)_F$
 relations among $b$-baryon decays and their numerical evaluation in section V. We conclude in section VI.
\section{Pentaquark spectrum in an effective Hamiltonian framework}
 We calculate the mass
spectrum of the pentaquarks by assuming that the underlying
structure is given by  $\bar{c}[cq][q^{\prime}q^{\prime \prime}]$, extending the
effective Hamiltonian proposed for the tetraquark spectroscopy by Maiani
\textit{et al.}~\cite{Maiani:2014aja}. The resulting Hamiltonian for pentaquarks
 is described in terms of the
constituent diquarks masses, $m_{[cq]}$, $m_{[q^{\prime}q^{\prime
    \prime}]}$, the spin-spin interactions between the quarks in each diquark shell, and the spin-orbit
and orbital angular momentum of the tetraquarks. To this are added 
 the charm quark mass $m_c$,  the spin-orbit and the orbital terms of the pentaquarks. Thus, in this picture,
there are two flux tubes, with the first stretched between the the two diquarks and the second between the
 tetraquark and the charm antiquark, with each string having its $L$ quantum number.   
\begin{equation}
H=H_{[\mathcal{Q}\mathcal{Q}^{\prime}]} + H_{\bar{c}[\mathcal{Q}\mathcal{Q}^\prime]} + H_{S_{\P} L_{\P}} + H_{L_{\P} L_{\P}}\label{main-Hamiltion} ,
\end{equation}
where the diquarks $[cq]$ and $[q^{\prime}q^{\prime \prime}]$ are denoted by  $\mathcal{Q}$ and $\mathcal{Q}
^{\prime}$ having masses $m_{\mathcal{Q}}$ and $m_{\mathcal{Q}^{\prime}}$,
respectively.  $L_{\P}$ and $S_{\P}$ are the orbital
angular momentum and the spin of the pentaquark state, and the quantities 
$A_{\mathcal{P}}$ and  $B_{\mathcal{P}}$, defined below,  parametrize the strength of their
spin-orbit  and orbital angular momentum couplings,  respectively. The individual terms in the
 Hamiltonian \eqref{main-Hamiltion} are
\begin{eqnarray}
\begin{array}{rclrclrcl}
H_{[\mathcal{Q}\mathcal{Q}^{\prime}]} &=&m_{\mathcal{Q}}+m_{\mathcal{Q}^{\prime}}+H_{SS}(\mathcal{Q}\mathcal{Q}^\prime) + H_{SL}(\mathcal{Q}\mathcal{Q}^\prime)+H_{LL}(\mathcal{Q}\mathcal{Q}^\prime),  &
H_{S_{\mathcal{P}}L_{\mathcal{P}}} &=&2A_{\mathcal{P}}(\mathbf{S}_{\mathcal{P}}\cdot
\mathbf{L_{\mathcal{P}}}), &
H_{L_{\mathcal{P}}L_{\mathcal{P}}}&=&B_{\mathcal{P}}\frac{L_{\mathcal{P}}(L_{\mathcal{P}}+1)}{2},
\end{array} 
\end{eqnarray}
and the other terms are
\begin{eqnarray}
H_{SS}(\mathcal{Q}\mathcal{Q}^\prime) &=& 2(\mathcal{K}_{cq})_{\bar{3}}(\mathbf{S}_{c}\cdot
\mathbf{S}_{q})+2(\mathcal{K}_{q^{\prime}q^{\prime \prime}})_{\bar{3}}(\mathbf{S}_{q^{\prime}}\cdot
\mathbf{S}_{q^{\prime \prime}}) \label{02}, \\
H_{\bar{c}[\mathcal{Q}\mathcal{Q}^{\prime}]}&=& m_c+2\,\mathcal{K}_{\bar{c}c}(\mathbf{S}_{\bar{c}}\cdot
\mathbf{S}_{c})+2\mathcal{K}_{\bar{c}q}(\mathbf{S}_{\bar{c}}\cdot
\mathbf{S}_{q}) +2\mathcal{K}_{\bar{c}q^{\prime}}(\mathbf{S}_{\bar{c}}\cdot
\mathbf{S}_{q^{\prime}})+2\mathcal{K}_{\bar{c}q^{\prime \prime}}(\mathbf{S}_{\bar{c}}\cdot
\mathbf{S}_{q^{\prime \prime}}), \label{02-1}
\end{eqnarray}
where 
  $(\mathcal{K}_{cq})_{\bar{3}%
}$ and $(\mathcal{K}_{q^{\prime}q^{\prime \prime}})_{\bar{3}}$ are the couplings corresponding to spin-spin interactions between 
the quarks inside the diquarks. Finally, the spin and orbital angular momentum couplings of the tetraquark are
\begin{equation}
H_{SL}(\Q\Q^\prime)  = 2 A_{\Q\Q^\prime} \mathbf{S}_{\Q \Q^\prime}\cdot \mathbf{L}_{\Q\Q^\prime}
,\;\;  H_{LL} = B_{\Q \Q^\prime} \frac{L_{\Q\Q^\prime}(L_{\Q\Q^\prime}+1)}{2}.
\label{03}
\end{equation}
In the earlier model proposed by Maiani \textit{et al.}~\cite{Maiani:2004vq}, in addition to the coupling between quarks inside the diquark, the couplings between the quarks of the two diquarks were also included. This 
extends the $H_{SS}(\Q\Q^\prime)$ part given above by  including four additional spin-spin terms,
\begin{equation}
H_{SS}(\Q\Q^\prime) = 2(\mathcal{K}_{cq})_{\bar{3}}(\mathbf{S}_{c}\cdot
\mathbf{S}_{q})+2(\mathcal{K}_{q^{\prime}q^{\prime \prime}})_{\bar{3}}(\mathbf{S}_{q^{\prime}}\cdot
\mathbf{S}_{q^{\prime \prime}})+2(\mathcal{K}_{cq^\prime})_{\bar{3}}(\mathbf{S}_{c}\cdot
\mathbf{S}_{q^\prime})+2(\mathcal{K}_{cq^{\prime \prime}})_{\bar{3}}(\mathbf{S}_{c}\cdot
\mathbf{S}_{q^{\prime \prime}})+2(\mathcal{K}_{qq^\prime})_{\bar{3}}(\mathbf{S}_{q}\cdot
\mathbf{S}_{q^\prime})+2(\mathcal{K}_{qq^{\prime \prime}})_{\bar{3}}(\mathbf{S}_{q}\cdot
\mathbf{S}_{q^{\prime \prime}}). \label{02-2}
\end{equation}
This then accounts for all possible spin-spin interactions; we have taken all the couplings to be positive.

The mass formula for the pentaquark state with the ground state tetraquark ($%
L_{{\mathcal{Q}}{\mathcal{Q}}^{\prime }}=0)$ can be written as 
\begin{equation}
M=M_{0}+\frac{B_{\mathcal{P}}}{2}L_{\mathcal{P}}(L_{\mathcal{P}}+1)+2A_{%
\mathcal{P}}\frac{J_{\mathcal{P}}(J_{\mathcal{P}}+1)-L_{\mathcal{P}}(L_{%
\mathcal{P}}+1)-S_{\mathcal{P}}(S_{\mathcal{P}}+1)}{2}+\Delta M
\label{mass-formula}
\end{equation}%
where $M_{0}=m_{\mathcal{Q}}+m_{\mathcal{Q}^{\prime }}+m_{c}$ and $\Delta M$ is the mass term
that arises from different spin-spin interactions. With the tetraquark in 
$L_{{\mathcal{Q}}{\mathcal{Q}}^{\prime }}=1$, one has to add the two terms
given above with their coefficients $A_{{\mathcal{Q}}{\mathcal{Q}}^{\prime }}
$ and $B_{{\mathcal{Q}}{\mathcal{Q}}^{\prime }}$. In this work, we restrict ourselves to the $S$-wave tetraquarks.

 For $L_{\mathcal{P}}=0$, we have classified the states in terms
of the diquarks spins, $S_{\mathcal{Q}}$ and $S_{\mathcal{Q}^{\prime }}$; the spin of anti-charm
quark is $S_{\bar{c}}=1/2$. There are
four $S$-wave pentaquark states for $J^{P}=\frac{3}{2}^{-}$ and a single
state with $J^{P}=\frac{5}{2}^{-}$. 
For $J^{P}=\frac{3}{2}^{-}$, we have the following states\footnote{For a similar classification in the diquark-triquark
picture, see\cite{Zhu:2015bba}.}:
\begin{eqnarray}
|0_{\mathcal{Q}},1_{\mathcal{Q}^{\prime }},\frac{1}{2}_{\bar{c}};\frac{3}{2}\rangle _{1} &=&%
\frac{1}{\sqrt{2}}[\left( \uparrow \right) _{c}\left( \downarrow \right)
_{q}-\left( \downarrow \right) _{c}\left( \uparrow \right) _{q}]\left(
\uparrow \right) _{q^{\prime }}\left( \uparrow \right) _{q^{\prime \prime
}}\left( \uparrow \right) _{\bar{c}}  \notag \\
|1_{\mathcal{Q}},0_{\mathcal{Q}^{\prime }},\frac{1}{2}_{\bar{c}};\frac{3}{2}\rangle _{2} &=&%
\frac{1}{\sqrt{2}}[\left( \uparrow \right) _{q^{\prime }}\left( \downarrow
\right) _{q^{\prime \prime }}-\left( \downarrow \right) _{q^{\prime }}\left(
\uparrow \right) _{q^{\prime \prime }}]\left( \uparrow \right) _{c}\left(
\uparrow \right) _{q}\left( \uparrow \right) _{\bar{c}}  \notag \\
|1_{\mathcal{Q}},1_{\mathcal{Q}^{\prime }},\frac{1}{2}_{\bar{c}};\frac{3}{2}\rangle _{3} &=&%
\frac{1}{\sqrt{6}}\left( \uparrow \right) _{c}\left( \uparrow \right)
_{q}\{2\left( \uparrow \right) _{q^{\prime }}\left( \uparrow \right)
_{q^{\prime \prime }}\left( \downarrow \right) _{\bar{c}}-[\left( \uparrow
\right) _{q^{\prime }}\left( \downarrow \right) _{q^{\prime \prime }}+\left(
\downarrow \right) _{q^{\prime }}\left( \uparrow \right) _{q^{\prime \prime
}}]\left( \uparrow \right) _{\bar{c}}\}  \notag \\
|1_{\mathcal{Q}},1_{\mathcal{Q}^{\prime }},\frac{1}{2}_{\bar{c}};\frac{3}{2}\rangle _{4} &=&%
\sqrt{\frac{3}{10}}[\left( \uparrow \right) _{c}\left( \downarrow \right)
_{q}+\left( \downarrow \right) _{c}\left( \uparrow \right) _{q}]\left(
\uparrow \right) _{q^{\prime }}\left( \uparrow \right) _{q^{\prime \prime
}}\left( \uparrow \right) _{\bar{c}}-\sqrt{\frac{2}{15}}\left( \uparrow
\right) _{c}\left( \uparrow \right) _{q}\{\left( \uparrow \right)
_{q^{\prime }}\left( \uparrow \right) _{q^{\prime \prime }}\left( \downarrow
\right) _{\bar{c}}  \notag \\
&&+[\left( \uparrow \right) _{q^{\prime }}\left( \downarrow \right)
_{q^{\prime \prime }}+\left( \downarrow \right) _{q^{\prime }}\left(
\uparrow \right) _{q^{\prime \prime }}]\left( \uparrow \right) _{\bar{c}}\},
\label{Spinj3by2}
\end{eqnarray}%
and the spin representation  corresponding  to $J^{P}=\frac{5}{2}^{-}$ state is\footnote{In the following, we shall
suppress the $\frac{1}{2}_{\bar{c}} $ quantum number, as it is the same for all the states discussed here.}

\begin{equation}
|1_{\mathcal{Q}},1_{\mathcal{Q}^{\prime }},\frac{1}{2}_{\bar{c}};\frac{5}{2}\rangle =\left(
\uparrow \right) _{c}\left( \uparrow \right) _{q}\left( \uparrow \right)
_{q^{\prime }}\left( \uparrow \right) _{q^{\prime \prime }}\left( \uparrow
\right) _{\bar{c}}.  \label{spinj5by2}
\end{equation}

Using the basis vector defined in Eq. (\ref{Spinj3by2}), the corresponding
mass splitting matrix $\Delta M$ may be obtained as%
\begin{equation}
\Delta M=\left( 
\begin{array}{cccc}
m _{11} & m _{12} & m _{13} & m _{14} \\ 
m _{21} & m _{22} & m _{23} & m _{24} \\ 
m _{31} & m _{32} & m _{33} & m _{34} \\ 
m _{41} & m _{42} & m _{43} & m _{44}%
\end{array}%
\right).  \label{splitting-matrix}
\end{equation}%
where, the different entries of the above matrix can be written in terms of spin-spin couplings as follow:
\begin{eqnarray}
m _{11} &=&\frac{1}{2}((\mathcal{K}%
_{q^{\prime }q^{\prime \prime }})_{\bar{3}}-3(\mathcal{K}_{cq})_{\bar{3}}+\mathcal{K}_{\bar{c}q^{\prime }}+%
\mathcal{K}_{\bar{c}q^{\prime \prime }}),  \notag \\
m _{12} &=&m _{21}=\frac{1}{2}((\mathcal{K}_{cq^{\prime }})_{%
\bar{3}}-(\mathcal{K}_{cq^{\prime \prime }})_{\bar{3}}-(\mathcal{K}%
_{qq^{\prime }})_{\bar{3}}+(\mathcal{K}_{qq^{\prime \prime }})_{\bar{3}}), 
\notag \\
m _{13} &=&m _{31}=\frac{1}{2\sqrt{3}}((\mathcal{K}%
_{cq^{\prime }})_{\bar{3}}+(\mathcal{K}_{cq^{\prime \prime }})_{%
\bar{3}}-(\mathcal{K}_{qq^{\prime }})_{\bar{3}}-(%
\mathcal{K}_{qq^{\prime \prime }})_{\bar{3}}+2\mathcal{K}_{\bar{c}q}-2\mathcal{%
K}_{\bar{c}c}), \notag \\
m _{14} &=&m _{41}=\frac{\sqrt{15}}{6}((\mathcal{K}_{cq^{\prime
}})_{\bar{3}}+(\mathcal{K}_{cq^{\prime \prime }})_{\bar{3}}-(\mathcal{K}%
_{qq^{\prime }})_{\bar{3}}-(\mathcal{K}_{qq^{\prime \prime }})_{\bar{3}}-%
\mathcal{K}_{\bar{c}q}+\mathcal{K}_{\bar{c}c}),  \notag \\
m _{22} &=&\frac{1}{2}((\mathcal{K}_{cq})_{\bar{3}}-3(\mathcal{K}_{q^{\prime }q^{\prime \prime }})_{%
\bar{3}}+\mathcal{K}_{\bar{c}q}+\mathcal{K}_{\bar{c}c}),  \label{matrix-entries} \\
m _{23} &=&m _{32}=\frac{1}{2\sqrt{3}}(-(\mathcal{K}%
_{cq^{\prime }})_{\bar{3}}+(\mathcal{K}_{cq^{\prime \prime }})_{%
\bar{3}}-(\mathcal{K}_{qq^{\prime }})_{\bar{3}}+(%
\mathcal{K}_{qq^{\prime \prime }})_{\bar{3}}),  \notag \\
m _{24} &=&m _{42}=\frac{\sqrt{15}}{6}(-(\mathcal{K}_{cq^{\prime
}})_{\bar{3}}+(\mathcal{K}_{cq^{\prime \prime }})_{\bar{3}}-(\mathcal{K}%
_{qq^{\prime }})_{\bar{3}}+(\mathcal{K}_{qq^{\prime \prime }})_{\bar{3}}), 
\notag \\
m _{33} &=&\frac{1}{6}(2((\mathcal{K}_{cq^{\prime }})_{\bar{3}}+(%
\mathcal{K}_{cq^{\prime \prime }})_{\bar{3}}+(\mathcal{K}_{qq^{\prime }})_{%
\bar{3}}+(\mathcal{K}_{qq^{\prime \prime }})_{\bar{3}})+3(\mathcal{K}_{cq})_{%
\bar{3}}+3(\mathcal{K}_{q^{\prime }q^{\prime \prime }})_{\bar{3}}  \notag \\
&&-\mathcal{K}_{\bar{c}q}-\mathcal{K}_{\bar{c}c}-6\mathcal{K}_{\bar{c}%
q^{\prime }}-6\mathcal{K}_{\bar{c}q^{\prime \prime }}),  \notag \\
m _{34} &=&m _{43}=\frac{\sqrt{5}}{3}(\mathcal{K}_{\bar{c}q}+%
\mathcal{K}_{\bar{c}c}-\frac{1}{2}(\mathcal{K}_{cq^{\prime }})_{\bar{3}}-%
\frac{1}{2}(\mathcal{K}_{cq^{\prime \prime }})_{\bar{3}}-\frac{1}{2}(%
\mathcal{K}_{qq^{\prime }})_{\bar{3}}-\frac{1}{2}(\mathcal{K}_{qq^{\prime
\prime }})_{\bar{3}}),  \notag \\
m _{44} &=&\frac{1}{6}(3(\mathcal{K}_{cq})_{\bar{3}}+3(\mathcal{K}%
_{q^{\prime }q^{\prime \prime }})_{\bar{3}}-2(\mathcal{K}_{cq^{\prime }})_{%
\bar{3}}-2(\mathcal{K}_{cq^{\prime \prime }})_{\bar{3}}-2(\mathcal{K}%
_{qq^{\prime }})_{\bar{3}}-2(\mathcal{K}_{qq^{\prime \prime }})_{\bar{3}} 
\notag \\
&&-2\mathcal{K}_{\bar{c}q}-2\mathcal{K}_{\bar{c}c}+3\mathcal{K}_{\bar{c}%
q^{\prime }}+3\mathcal{K}_{\bar{c}q^{\prime \prime }}).  \notag
\end{eqnarray}%
The couplings $(\mathcal{K}_{cq^{\prime }})_{\bar{3}}$, $(\mathcal{K}%
_{cq^{\prime \prime }})_{\bar{3}}$, $(\mathcal{K}_{qq^{\prime }})_{\bar{3}}$%
, $(\mathcal{K}_{qq^{\prime \prime }})_{\bar{3}}$ given in the above expressions correspond to the
spin-spin interactions between the quarks of the two diquarks in Model I 
\cite{Maiani:2004vq}, which are set to zero in the later version, Model II  
\cite{Maiani:2014aja}.

The masses for the four $S$-wave pentaquark states with
$J^{P}=\frac{3}{2}^{-}$ and a single state with $J^{P}=\frac{5}{2}^{-}$ are given in Table \ref{Table I}, where we
 label the states as $\mathcal {P}_{X_i}$. The corresponding 
five $P$-wave pentaquark states with $L_{\mathcal{P}}=1$ and $J^{P}=\frac{5}{2}^{+}$  are labeled as $\mathcal {P}_{Y_i}$ in
Table \ref{Table I}.
\footnotesize
\begin{table}[tb]
\caption{\sf $S$ ($P$)- wave pentaquark states $\mathcal{P}_{X_{i}}$ ($\mathcal{P}_{Y_{i}}$)
and their spin- and orbital angular momentum quantum numbers. The subscripts  $\mathcal{Q}$ and 
$\mathcal{Q}^{\prime}$ represent the heavy $[cq]$ and light $[q^{\prime}
q^{\prime \prime}]$ diquarks, respectively. In the expressions for the masses of the
$\mathcal{P}_{Y_{i}}$ states, the terms $M_{\mathcal{P}_{X_{i}}}= M_0 + \Delta M_i$ with $i=1,...,5$.}
\label{Table I}
%\begin{center}
\begin{tabular}{|l|l|l||l|l|l|l|l|l|l|}
\hline
Label  &
% \small
 $|S_{\mathcal{Q}},S_{\mathcal{Q}^{\prime }};L_{\mathcal{P}},J^{P}\rangle _{i}
$ & Mass & Label & $|S_{\mathcal{Q}},S_{\mathcal{Q}^{\prime }};L_{\mathcal{P}%
},J^{P}\rangle _{i}$ & Mass \\ \hline
$\mathcal{P}_{X_{1}}$ & $|0_{\mathcal{Q}},1_{\mathcal{Q}^{\prime }},0;%
\frac{3}{2}^{-}\rangle _{1}$ & $M_{0}+\Delta M_{1}$ & $\mathcal{P}_{Y_{1}}$
& $|0_{\mathcal{Q}},1_{\mathcal{Q}^{\prime }},1;\frac{5}{2}^{+}\rangle _{1}
$ & $M_{\mathcal{P}_{X_{1}}}+3A_{\mathcal{P}}+B_{\mathcal{P}}$ \\ \hline
$\mathcal{P}_{X_{2}}$ & $|1_{\mathcal{Q}},0_{\mathcal{Q}^{\prime }},0;%
\frac{3}{2}^{-}\rangle _{2}$ & $M_{0}+\Delta M_{2}$ & $\mathcal{P}_{Y_{2}}$
& $|1_{\mathcal{Q}},0_{\mathcal{Q}^{\prime }},1;\frac{5}{2}^{+}\rangle _{2}
$ & $M_{\mathcal{P}_{X_{2}}}+3A_{\mathcal{P}}+B_{\mathcal{P}}$ \\ \hline
$\mathcal{P}_{X_{3}}$ & $|1_{\mathcal{Q}},1_{\mathcal{Q}^{\prime }},0;%
\frac{3}{2}^{-}\rangle _{3}$ & $M_{0}+\Delta M_{3}$ & $\mathcal{P}_{Y_{3}}$
& $|1_{\mathcal{Q}},1_{\mathcal{Q}^{\prime }},1;\frac{5}{2}^{+}\rangle _{3}
$ & $M_{\mathcal{P}_{X_{3}}}+3A_{\mathcal{P}}+B_{\mathcal{P}}$ \\ \hline
$\mathcal{P}_{X_{4}}$ & $|1_{\mathcal{Q}},1_{\mathcal{Q}^{\prime }},0;%
\frac{3}{2}^{-}\rangle _{4}$ & $M_{0}+\Delta M_{4}$ & $\mathcal{P}_{Y_{4}}$
& $|1_{\mathcal{Q}},1_{\mathcal{Q}^{\prime }},1;\frac{5}{2}^{+}\rangle _{4}
$ & $M_{\mathcal{P}_{X_{4}}}+3A_{\mathcal{P}}+B_{\mathcal{P}}$ \\ \hline
$\mathcal{P}_{X_{5}}$ & $|1_{\mathcal{Q}},1_{\mathcal{Q}^{\prime }},0;%
\frac{5}{2}^{-}\rangle _{5}$ & $M_{0}+\Delta M_{5}$ & $\mathcal{P}_{Y_{5}}$ & $|1_{\mathcal{Q}},1_{\mathcal{Q}^{\prime }},%
\frac{1}{2}_{\bar{c}},1;\frac{5}{2}^{+}\rangle _{5}$ & $M_{\mathcal{P}%
_{X_{5}}}-2A_{\mathcal{P}}+B_{\mathcal{P}}$ \\ \hline\hline
\end{tabular}%
%\end{center}
\end{table}
\normalsize
Here $\Delta M_{i}$  ($i$ runs from $1$ to $4$) are the mass splitting terms that arise after the
diagonalizing the $4 \times 4$ matrix (\ref{splitting-matrix}), whereas $\Delta M_{5}$ is given by
\begin{equation}
\Delta M_{5} = \frac{1}{2}(\mathcal{K}_{\bar{c}q}+\mathcal{K}_{\bar{c}c}+\mathcal{K}_{\bar{c}%
q^{\prime }}+\mathcal{K}_{\bar{c}q^{\prime \prime }}+(\mathcal{K}_{cq^{\prime }})_{\bar{3}}+(\mathcal{K}%
_{cq^{\prime \prime }})_{\bar{3}}+(\mathcal{K}_{qq^{\prime }})_{\bar{3}}%
+(\mathcal{K}_{qq^{\prime \prime }})_{\bar{3}}+(\mathcal{K}_{cq})_{\bar{3}}+(\mathcal{K}%
_{q^\prime q^{\prime \prime }})_{\bar{3}}).
\end{equation}
%%%%%%%%%%%%%%%%%%%%%%%%%%%%%%%%%%%%%%%%%%%%%%%%%%%%%%%%%%%%%%%%%%%%
%
\section{ $\boldsymbol{S}$- and $\boldsymbol{P}$-wave pentaquark spectrum with a $\boldsymbol{c\bar{c}}$ pair and three light quarks}
To evaluate numerically the  pentaquark mass spectrum, we  use the values of  the input
parameters extracted from the hidden $c \bar{c}$ states $(X, Y, Z)$~\cite{Maiani:2004vq}, where the  coupling of the 
heavy-light diquark $(\mathcal{K}_{cq})_{\bar{3}}$ $(q = u\; , d)$
and its mass $m_{\mathcal{Q}}$ are estimated  to be $110$ MeV and $1980$ MeV,
respectively. Corresponding values for the other diquarks couplings $(\kappa_{ij})_{\bar{3}}$ are summarized in Table \ref{Table II-bar}.  As we are working with the pentaquark states in the hidden charm sector,  we fixed  the value of the spin-orbit coupling
from the corresponding coupling in the hidden charm tetraquark sector, yielding   $A_{\mathcal{P}}=52$ MeV~\cite{Maiani:2014aja}. 

\begin{table}[tb]
\caption{\sf Constituent quark masses derived from the $L=0$ mesons and baryons.}
\label{Table II}
\begin{center}
\begin{tabular}{|l|l|l|l|l|l}
\hline
Constituent quark & $q$ & $s$ & $c$ & $b$ \\ \hline
Mass (MeV) [Mesons] & $305$ & $490$ & $1670$ & $5008$ \\ \hline
Mass (MeV) [Baryons] & $362$ & $546$ & $1721$ &$5050$ \\ \hline\hline
\end{tabular}%
\end{center}
\end{table}
%%%

For the pentaquarks, consisting of an anti-charm quark and  a tetraquark $[\mathcal{Q}\mathcal{Q}^{\prime}]$,
various spin-spin couplings  $\left( \mathcal{K}_{ij}\right) _{0}$  and $(\mathcal{K}_{ij})_{\bar{3}}$ enter in the
mass formulae,  whose values  are summarized in Table \ref{Table II-bar}.  Fixing these couplings, we are left with  one unknown parameter  $B_{\mathcal{P}}$,  involving the pentaquark orbital angular momentum. To estimate this,
  we identify the state  $|1_{\mathcal{Q}},0_{\mathcal{Q}^{\prime }},1;\frac{5}{2}^{+}\rangle _{2}$
  as the pentaquark state $\frac{5}{2}^+$ having a mass $4450$ MeV  and $L_{\mathcal{P}}=1$. This yields 
  $160$ MeV and  $220$ MeV for  $B_{\mathcal{P}}$  in the type I and type II models, respectively. 

\begin{table}[tb]
\caption{Spin-Spin couplings for quark-antiquark and the quark-quark pairs
in the color
singlet and triplet states from the known mesons.}
\label{Table II-bar}
\begin{center}
\begin{tabular}{|l|l|l|l|l|l|l||l|l|l|l|l|l|}
\hline
quark-antiquark & $q\bar{q}$ & $s\bar{q}$ & $s\bar{s}$ & $c\bar{q}$ &
$c\bar{s}$ & $c\bar{c}$ &
quark-quark & $qq$ & $sq$ & $ss$ & $cq$ & $cs$  \\ \hline
$\left( \mathcal{K}_{ij}\right) _{0}$(MeV) & $318$ & $200$ & $129$ &
$70$ & $72$ & $59$&
$(\kappa_{ij})_{\bar{3}}$ (MeV) & $103$ & $64$ & $126$ & $22$ &$25$ \\
\hline\hline
\end{tabular}%
\end{center}
\end{table}

\begin{table}[tb]
\caption{\sf Quark contents  (with $q=u$ or $d$) and the corresponding flavor labels $c_i$ ($i=1,...,5$) used in the text for the pentaquark states.}
\label{Table III}
%\begin{center}
\begin{tabular}{|l|l|l|l|l|l|}
\hline
Quark contents &$\bar{c}[cq][qq]$ &$\bar{c}[cq][sq]$&
$\bar{c}[cs][qq]$ &$\bar{c}[cs][sq]$ &$\bar{c}[cq][ss]$ \\ 
\hline
Label &$\quad c_1$ &$\quad c_2$&$\quad c_3$ &$\quad c_4$ &$\quad c_5$\\
\hline\hline
\end{tabular}%
%\end{center}
\end{table}

\begin{figure}[t]
\begin{tabular}{ccc}
\hspace{0.2cm}($\mathbf{a}$)&\hspace{0.4cm}($\mathbf{b}$)&\hspace{0.6cm}($\mathbf{c}$)\\
\includegraphics[scale=0.65]{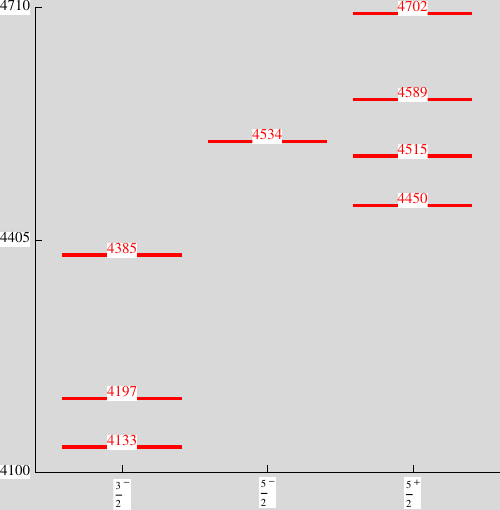}  \ \ \
&  \ \ \  \includegraphics[scale=0.65]{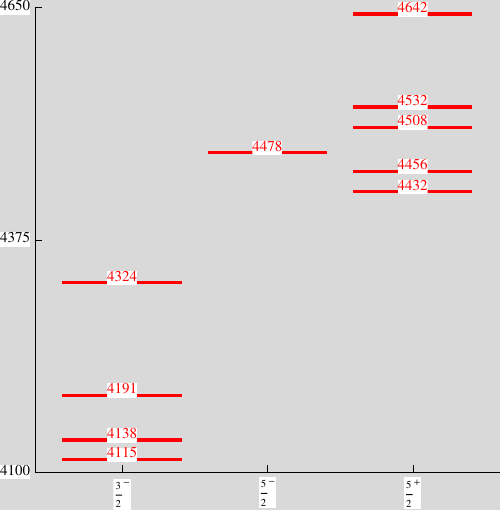}\ \ \
&  \ \ \  \includegraphics[scale=0.65]{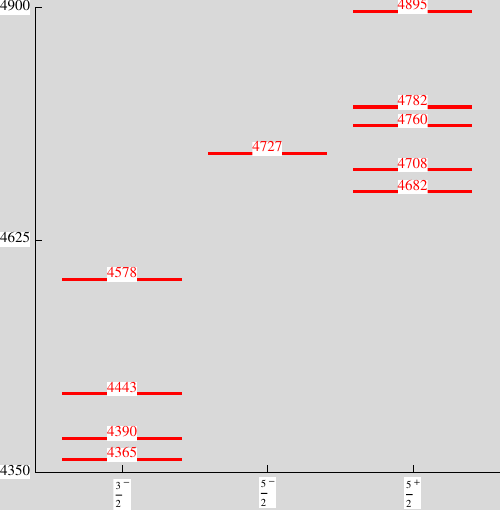}\end{tabular}
\caption{\sf Mass Spectrum (in MeV) of the lowest $S$- and $P$-wave pentaquark states in the
  diquark-diquark-antiquark picture for the charmonium sector using the
  tetraquark type-I model. Here,  (a) $c_1$, (b) $c_2$ and (c)  $c_3$
 are the labels for the pentaquarks with their quark contents given in Table \ref{Table III}.}
\label{Fig-Spectrum1}
\end{figure}

\begin{figure}[t]
\begin{tabular}{ccc}
\hspace{0.5cm}($\mathbf{a}$)&\hspace{1.0cm}($\mathbf{b}$)\\
\includegraphics[scale=0.60]{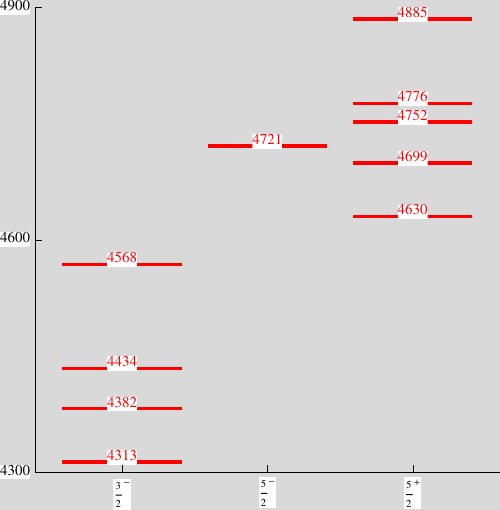}  \ \ \
&  \ \ \  \includegraphics[scale=0.60]{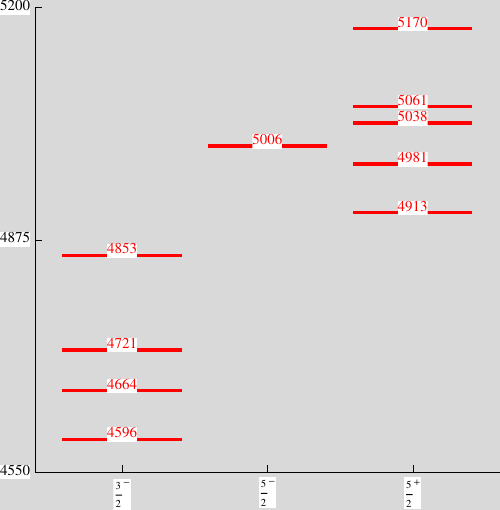}\end{tabular}
\caption{\sf Mass Spectrum (in MeV) of the lowest $S$-and $P$-wave pentaquark states in the
  diquark-diquark-antiquark picture for the charmonium sector using the
  tetraquark type-I model. Here,
(a) $c_4$ and (b) $c_5$ are the labels for the pentaquarks with their quark contents given in Table \ref{Table III}.} 
\label{Fig-Spectrum2}
\end{figure}

\begin{figure}[t]
\begin{tabular}{ccc}
\hspace{0.2cm}($\mathbf{a}$)&\hspace{0.4cm}($\mathbf{b}$)&\hspace{0.6cm}($\mathbf{c}$)\\
\includegraphics[scale=0.65]{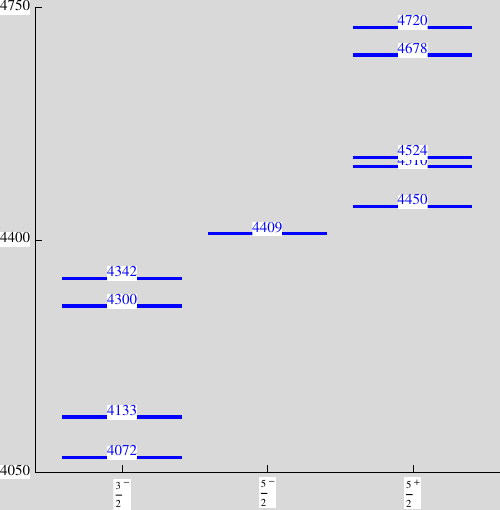}  \ \ \
&  \ \ \  \includegraphics[scale=0.65]{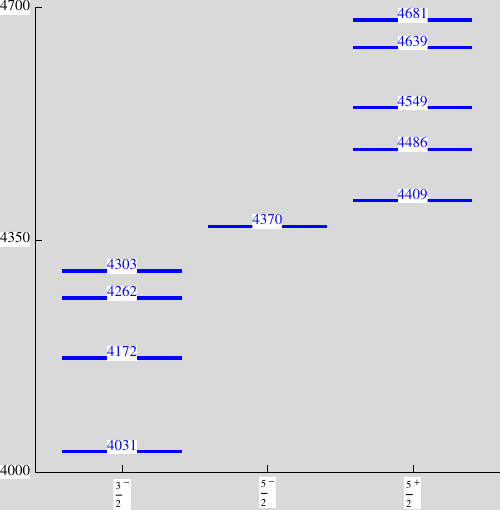}\ \ \
&  \ \ \  \includegraphics[scale=0.65]{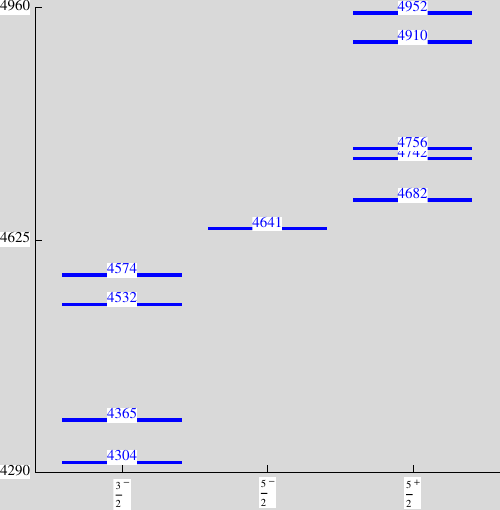}\end{tabular}
\caption{\sf Mass Spectrum (in MeV)  of the lowest $S$-and $P$-wave pentaquark states in
the diquark-diquark-antiquark picture for the charmonium sector using
 the tetraquark type-II model. Here, (a) $c_1$, (b) $c_2$ and (c)  $c_3$
 are labels for the pentaquarks with their quark contents given in Table \ref{Table III}.}
  \label{Fig-Spectrum1-II}
\end{figure}

\begin{figure}[t]
\begin{tabular}{ccc}
\hspace{0.5cm}($\mathbf{a}$)&\hspace{1.0cm}($\mathbf{b}$)\\
\includegraphics[scale=0.60]{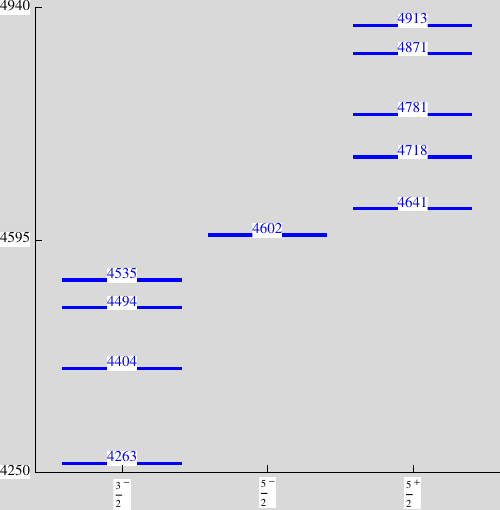}  \ \ \
&  \ \ \  \includegraphics[scale=0.60]{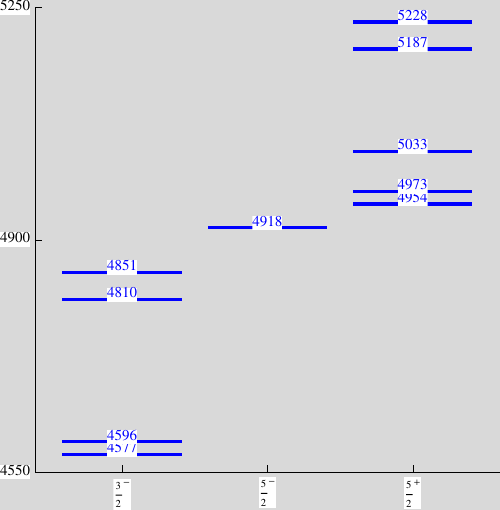}\end{tabular}
\caption{\sf  Mass Spectrum (in MeV)  of the lowest $S$-and $P$-wave pentaquark states in
the diquark-diquark-antiquark picture for the charmonium sector using
 the tetraquark type-II model. Here, 
(a) $c_4$ and (b) $c_5$ are labels for the pentaquarks with their quark contents given in Table \ref{Table III}. } 
\label{Fig-Spectrum2-II}
\end{figure}

With these input values, we have calculated the mass spectrum for the different pentaquark states
which have a $c \bar{c}$ pair and three light quarks belonging to $SU(3)_F$. The results are
 shown in Figs. \ref{Fig-Spectrum1} and  \ref{Fig-Spectrum2}, for the parameters in Model I, and in Figs.
  \ref{Fig-Spectrum1-II} and \ref{Fig-Spectrum2-II}, for Model II. The labels $c_i$ used in these figures 
 specify the quark flavor content of the pentaquark states, explicitly given in Table \ref{Table III}.
 The corresponding numerical values for the masses and their errors in both the type I and type II models are given in Tables \ref{TableIV-1} and \ref{TableIV-2} for $\mathcal{P}_{{X}_i}$ and $\mathcal{P}_{{Y}_i}$ states, respectively. 
Obviously, the errors are of parametric origin, assuming the effective Hamiltonian, and not from the assumed
form of the Hamiltonian. 
The errors shown arise from the uncertainties in the spin-spin couplings of the quarks inside the diquarks, the masses of the diquarks,  and also the couplings of the anti-charm quark with the quarks in the colored tetraquark states. To estimate these errors, we have used the couplings from the hidden  charm and the hidden bottom tetraquark states.
As an example, by using the states $Z_b^{+}(10610)$ and $Z_b^{+}(10650)$, the heavy-light diquark 
coupling  $(\mathcal{K}_{bq})_{\bar{3}}$ is estimated  to be $23$ MeV \cite{Ali:2014dva}. With the
relation  $(\mathcal{K}_{q^{\prime}q^{\prime \prime}})_{\bar{3}}=(\mathcal{K}_{bq})_{\bar{3}} \times m_{b}/m_{q^{\prime}}$, and a similar one involving the coupling  $(\mathcal{K}_{bc})_{\bar{3}}$, we can get 
two different  estimates of the quark-quark couplings inside the light diquark. 
 With this,  and the uncertainties on the masses of the diquarks, we have calculated the errors in the masses of the pentaquark states which are given in the Tables \ref{TableIV-1} and \ref{TableIV-2}. Typical parametric errors  in this way are about $\pm$  50 MeV.

Entries in Table  \ref{TableIV-1}   in the first row,
corresponding to the flavor label $c_1$ (representing the quark content $\bar{c}[cq][qq]$ with $q = u\; ,\ d$)
show that indeed a state $\mathcal{P}_{{X}_4}$  is predicted with the mass 4385 (4342) MeV in the type I (II) model, having the  quantum numbers  $|1_{\mathcal{Q}},1_{\mathcal{Q}^{\prime }},0;\frac{3}{2}^{-}\rangle $.  This agrees with the mass of the observed state $P^+_c(4380)$, and the spin assignment matches with the proposal by Maiani \textit{et al.} \cite{Maiani:2015vwa}. Likewise, identifying the state $P^+_c(4550)$, having $J^P=\frac{5}{2}^+$ with the state $\mathcal{P}_{{Y}_2}$
in the first row of Table \ref{TableIV-2} having the quantum numbers $|1_{\mathcal{Q}},0_{\mathcal{Q}^{\prime }},1;\frac{5}{2}^{+}\rangle $,
also fits well with the one given by Maiani \textit{et  al.} \cite{Maiani:2015vwa}. The point where we differ is that
with the heavy quark symmetry imposed, the state  $\mathcal{P}_{{X}_4}$ is unlikely to be produced in
$\Lambda_b$ decays, as it has the wrong light-diquark spin. On the other hand, we do have a 
  lower mass state $\mathcal{P}_{{X}_2}$, having the correct flavor and spin quantum numbers
  $|1_{Q},0_{Q^{\prime }},0;\frac{3}{2}^{-}\rangle$,  with  a mass of about
  4130 MeV, which we expect to be produced in $\Lambda_b$ decays. One could argue that our mass estimates
  following from the assumed effective Hamiltonian are in error by a larger amount than what we quote.
  However, as already stated, the mass difference
  between the $J^P=\frac{5}{2}^+$ and $J^P=\frac{3}{2}^-$ pentaquarks, having the right quantum numbers 
  $|1_{\mathcal{Q}},0_{\mathcal{Q}^{\prime }},1;\frac{5}{2}^{+}\rangle $ and $|1_{\mathcal{Q}},0_{\mathcal{Q}^{\prime }},0;\frac{3}{2}^{-}\rangle$
   is expected to be around 340 MeV, yielding a mass for
  the  lower-mass $J^P=\frac{3}{2}^-$ pentaquark state of about 4110 MeV. The two estimates are compatible with
  each other, and we advocate to search for this state in the LHCb data. Among the ten states listed in the
  first rows of Tables \ref{TableIV-1} and \ref{TableIV-2}, only the ones called  $\mathcal{P}_{{X}_2}$ and 
 $\mathcal{P}_{{Y}_2}$ are allowed as the $\Lambda_b$ decay products. 

%
%%
%%%%%%%%%%%%%%%%%%%%% Spectrums %%%%%%%%%%%%%%%%%%%%%%%%
%%
%


%
%%%%%%%%%%%%%%%%%%%%% Spectrums in tables %%%%%%%%%%%%%%%%%%%%%%%%
%\
\begin{table*}[tbp]
\caption{\sf Masses of the hidden charm $S$-wave pentaquark states $\mathcal{P}_{X_i}$ (in MeV) formed through different diquark-diquark-anti-charm quark combinations in type I and type II models of tetraquarks. The quoted errors are obtained from the uncertainties in the input parameters in the effective
Hamiltonian.The masses given in the parentheses are for the input values taken from the type II model. }
\label{TableIV-1}
%\begin{center}
\begin{tabular}{|l|l|l|l|l|l|}
\hline
$\mathcal{P}_{X_{i}} \quad$& $\mathcal{P}_{X_{1}}$ & $\mathcal{P}_{X_{2}}$
& $\mathcal{P}_{X_{3}}$ &$\mathcal{P}_{X_{4}} $  &$\mathcal{P}_{X_{5}}
$ \\ 
\hline
$c_1$ & $4133\pm55$ $(4072 \pm 40)$ & $4133\pm55$ $(4133 \pm 55)$ &
$4197\pm55$ $(4300 \pm 40)$ & $4385\pm55$ $(4342 \pm 40)$ & $4534\pm55$ $(4409 \pm 40)$   \\ 
\hline
$c_2$ & $4115\pm58$ $(4031 \pm 43)$ & $4138\pm 47$ $(4172 \pm 47)$ & $ 4191\pm 53$ $(4262 \pm 43)$ & $4324\pm 47$ $(4303 \pm 43)$ & $4478\pm 47$ $(4370 \pm 43)$  \\ 
\hline
$c_3$ & $4365\pm 55$ $(4304 \pm 55)$  & $ 4390\pm 42$ $(4365 \pm 40)$ &$ 4443\pm 49$ $(4532\pm 40)$  & $4578\pm 43$ $(4574\pm 40)$ & $4727\pm 42$ $(4641 \pm 40)$ \\ 
\hline
$c_4$ & $4313\pm 47$ $(4263 \pm 43)$ & $ 4382\pm 45$ $(4404 \pm 47)$ & $ 4434\pm 51$ $(4494 \pm 43)$ & $4568\pm 46$ $(4535 \pm 43)$ & $4721\pm 45$ $(4602 \pm 43)$\\ 
\hline
$c_5$ & $4596\pm 47$ $(4577  \pm 43)$ & $ 4664\pm 46$ $(4596 \pm 47)$ & $4721\pm 51$ $(4810 \pm 43)$ & $4853\pm 46$ $(4851 \pm 43)$ & $5006\pm 45$ $(4918 \pm 47)$\\ \hline \hline%
\end{tabular}%\
%\end{center}
\end{table*}%
%%%%%%%%%%
%%%%%%%%%%%%%%%%%%%%%%%%%%%%%%%%%%%%%%%%%%%%%%%%%%%%%%%%%%%%%%%%%%%%%%%%%%%%%%
%\
\begin{table*}[tbp]
\caption{\sf Masses of the hidden charm $P$-wave pentaquark states $\mathcal{P}_{Y_i}$ (in MeV) formed through different diquark-diquark-anti-charm quark combinations in type I and type II models of tetraquarks. The quoted errors are obtained from the uncertainties in the input parameters in the effective
Hamiltonian.The masses given in the parentheses are for the input values taken from the type II model.}
\label{TableIV-2}
%\begin{center}
\begin{tabular}{|l|l|l|l|l|l|}
\hline
$\mathcal{P}_{Y_{i}} \quad$& $\mathcal{P}_{Y_{1}}$ & $\mathcal{P}_{Y_{2}}$
& $\mathcal{P}_{Y_{3}}$ &$\mathcal{P}_{Y_{4}} $  &$\mathcal{P}_{Y_{5}}$ \\ 
\hline
$c_1$ & $4450\pm 57 $ $(4450 \pm 44)$ & $4450\pm 57$ $(4510 \pm 57)$ & $4515\pm 57 $ $(4678 \pm 44)$ & $4702\pm 58$ $(4720 \pm 44)$ & $4589\pm 56$ $(4524 \pm 41)$  \\ 
\hline
$c_2$ & $ 4432\pm61$ $(4409 \pm 47)$ & $4456\pm 50$ $(4549 \pm 51)$ & $4508\pm 56$ $(4639 \pm 47)$ & $4642\pm 50$ $(4681 \pm 47)$ & $4532\pm 48$ $(4486 \pm 45)$  \\ 
\hline
$c_3$ & $ 4682\pm 57$ $(4682  \pm 44)$ & $4708\pm 46$ $(4742 \pm 57)$ & $4760\pm 52$ $(4910  \pm 44)$ & $4895\pm 47$ $(4952  \pm 44)$ & $4782\pm 44$ $(4756  \pm 41)$  \\ 
\hline
$c_4$ & $4603\pm 51$ $(4641 \pm 47)$ & $ 4699\pm 49$ $(4781  \pm 51)$ & $ 4752\pm 54$ $(4871 \pm 47)$  &$4885\pm 49$ $(4913 \pm 47)$ & $4776\pm 47$ $(4718 \pm 45)$ \\ 
\hline
$c_5$ & $4913\pm 51$ $(4954 \pm 47)$ & $ 4981\pm 49$ $(4973 \pm 51)$ & $5038\pm 54$ $(5187 \pm 47)$ & $5170\pm 49$ $(5228 \pm 47)$ & $5061\pm 47$ $(5033 \pm 47)$\\ 
\hline\hline%
\end{tabular}%
%\end{center}
\end{table*}
%%%%%%%%%%%%%%%%%%%%%%%%%%%%%%%%%%%%%%%%%%%%%%%%%%%%%%%%%%%%%%%%%%%%%%%%
%
\section{Weak Decays of the $\boldsymbol{b}$-baryons into pentaquark states}
In the previous sections, we have worked out the spectroscopy of the hidden charm  $S$- and $P$-wave pentaquarks with their
flavor structure displayed in Table \ref{Table III}. There are fifty such states anticipated
 having  masses estimated to lie in the range 4100 - 5100 MeV. They can, in principle, be produced in prompt production processes at the LHC. In practice, only a few have a chance to be detected
in experiments due to their significantly lower cross sections, as compared to the corresponding ones for tetraquarks,
  and the formidable experimental
background. As most of the multiquark states ($X,Y,Z,P_c$) have been observed in the decays of the
$B$ mesons and $\Lambda_b$,  we anticipate that some of the
pentaquark states discussed earlier can be  measured in $b$-baryon decays. However, 
 only those states obeying the flavor constraints
following from the weak Hamiltonian and having the  internal spin quantum numbers compatible with the
heavy quark symmetry will actually be produced in $b$-baryon decays.
The state $P_c^+(4450)$ discovered by the LHCb fits all the criterion to be a pentaquark
state, and we expect another $J^P=3/2^-$ pentaquark around 4110 MeV, but  a number of
other pentaquarks are anticipated  in $b$-baryon decays. Heavy quark symmetry reduces the otherwise allowed decay topologies (diagrams) and simplifies the discussion of various contributions.  $SU(3)_F$ symmetry can be invoked 
to relate some of these allowed decay rates, following \cite{Cheng-Chua, Li-He-He}, resulting in a number of
predictions to be tested in the future.

The decay amplitudes of interest can be generically written as%
\begin{equation}
\mathcal{A} = \left\langle \mathcal{PM}\left\vert H^W_{\text{eff}}\right\vert 
\mathcal{B}\right\rangle, \label{production-amplitude}
\end{equation}
where, $ H^W_{\text{eff}}$ is the effective weak Hamiltonian inducing the Cabibbo-allowed
$\Delta I=0, \Delta S =-1$ transition $b \to c \bar{c} s$, and the Cabibbo-suppressed $\Delta S =0$ transition $b \to c \bar{c} d$.
\begin{equation}
 H^W_{\text{eff}} = \frac{4 G_F}{\sqrt{2}} \left[ V_{cb} V_{cq}^* ( c_1 O_1^{(q)} +   c_2  O_2^{(q)} ) \right].
\label{weak-hamiltonian}
\end{equation}

Here, $G_F$ is the Fermi coupling constant, $V_{ij}$ are the CKM matrix elements, and 
$c_i$ are the Wilson coefficients of  the operators $O_1^{(q)}$ ($q=d,\; s  $), defined as
\begin{equation}
  O_1^{(q)}=  (\bar{q}_\alpha c_\beta)_{V-A} (\bar{c}_\alpha b_\beta)_{V-A};\;\;\;
  O_2^{(q)}=  (\bar{q}_\alpha c_\alpha)_{V-A} (\bar{c}_\beta b_\beta)_{V-A},
\label{tree-operators}
\end{equation}
where $\alpha$ and $\beta$ are $SU(3)$ color indices, and $V-A= 1 -\gamma_5$ reflects that the
charged currents are left-handed.

 In (\ref{production-amplitude}), $\mathcal{B}$ is a flavor
anti-triplet $b$-baryon with the light-quark spin $j^P=0^+$ 
(the  flavor sextet $b$-baryons with $j^P=0^+$ are denoted by $\mathcal{C}$),\\

$\qquad \qquad \qquad \qquad  \mathcal{B}_{ij}\left(\bar{3}\right) =\left( 
\begin{array}{ccc}
0 & \Lambda _{b}^{0} & \Xi _{b}^{0} \\ 
-\Lambda _{b}^{0} & 0 & \Xi _{b}^{-} \\ 
-\Xi _{b}^{0} & -\Xi _{b}^{-} & 0%
\end{array} \right) $,
$\qquad \mathcal{C}_{ij}\left( 6\right) =\left( 
\begin{array}{ccc}
\Sigma _{b}^{+} & \frac{\Sigma _{b}^{0}}{\sqrt{2}} & \frac{\Xi _{b}^{\prime
0}}{\sqrt{2}} \\ 
\frac{\Sigma _{b}^{0}}{\sqrt{2}} & \Sigma _{b}^{-} & \frac{\Xi _{b}^{\prime
-}}{\sqrt{2}} \\ 
\frac{\Xi _{b}^{\prime 0}}{\sqrt{2}} & \frac{\Xi _{b}^{\prime -}}{\sqrt{2}}
& \Omega _{b}^{-}%
\end{array}%
\right). $\\
 $\mathcal{M}$ is a light pseudoscalar meson in the $SU(3)_F$ octet,   and 
 $\mathcal{P}$ is the final state pentaquark state\\

$\qquad \qquad \mathcal{M}_{i}^{j}=\left( 
\begin{array}{ccc}
\frac{\pi ^{0}}{\sqrt{2}}+\frac{\eta _{8}}{\sqrt{6}} & \pi ^{+} & K^{+} \\ 
\pi ^{-} & -\frac{\pi ^{0}}{\sqrt{2}}+\frac{\eta _{8}}{\sqrt{6}} & K^{0} \\ 
K^{-} & \bar{K}^{0} & -\frac{2\eta _{8}}{\sqrt{6}}%
\end{array}%
\right)$,
$\qquad \mathcal{P}_{i}^{j}\left( J^{P}\right) =\left( 
\begin{array}{ccc}
\frac{P_{\Sigma ^{0}}}{\sqrt{2}}+\frac{P_{\Lambda }}{\sqrt{6}} & P_{\Sigma
^{+}} & P_{p} \\ 
P_{\Sigma ^{-}} & -\frac{P_{\Sigma ^{0}}}{\sqrt{2}}+\frac{P_{\Lambda }}{%
\sqrt{6}} & P_{n} \\ 
P_{\Xi ^{-}} & P_{\Xi ^{0}} & -\frac{P_{\Lambda }}{\sqrt{6}}%
\end{array}%
\right) .$\\

\noindent
 The  hidden-charm pentaquark 
states are classified by  their $J^P$ quantum numbers and their light-quark contents, usually specified by the
corresponding light baryons indicated by a
 subscript. Thus, the two states $P_c^+(J^P = \frac{3}{2}^-)$ and $P_c^+(J^P = \frac{5}{2}^+)$,
 having the flavor content $c\bar{c} uud$, are denoted here
by the components $P_p(J^P = \frac{3}{2}^- )$ and  $P_p(J^P = \frac{5}{2}^+)$. For the $J^P=3/2^-$, also a
decuplet is present $\mathcal{P}_{i j k}$ (symmetric in all indices) and they can be listed as:  $\mathcal{P}_{1 1 1} = P_{\Delta^{++}_{10}}\; , \mathcal{P}_{1 1 2} = P_{\Delta^{+}_{10}}/\sqrt{3}\; , \mathcal{P}_{1 2 2} = P_{\Delta^{0}_{10}}/\sqrt{3}\; , \mathcal{P}_{2 2 2} = P_{\Delta^{-}_{10}}\;,  \mathcal{P}_{1 1 3} = P_{\Sigma^{+}_{10}}/\sqrt{3}\; , \mathcal{P}_{1 2 3} = P_{\Sigma^{0}_{10}}/\sqrt{6}\; , \mathcal{P}_{2 2 3} = P_{\Sigma^{-}_{10}}/\sqrt{3}\; , \mathcal{P}_{1 3 3} = P_{\Xi^{0}_{10}}/\sqrt{3}\; , \mathcal{P}_{2 3 3} = P_{\Xi^{-}_{10}}/\sqrt{3}\;$ and $\mathcal{P}_{3 3 3} = P_{\Omega^{-}_{10}}$ \cite{Li-He-He}.

\noindent
With these definitions, the  tree amplitudes for the anti-triplet $b$-baryon decays 
into an octet pentaquark and a pseudoscalar meson, denoted by $\mathcal{A}^{J}_{t8}$, 
 can be decomposed  as \cite{Li-He-He}%
\begin{eqnarray*}
\mathcal{A}^{J}_{t8}\left( q\right)  &=&T^{J}_{1}\left\langle \mathcal{P}_{l}^{k}%
\mathcal{M}_{k}^{l}\left\vert H\left( \bar{3}\right) ^{i}\right\vert 
\mathcal{B}_{i^{\prime }i^{\prime \prime }}\right\rangle \varepsilon
^{ii^{\prime }i^{\prime \prime }}+T^{J}_{2}\left\langle \mathcal{P}_{j}^{k}%
\mathcal{M}_{k}^{i}\left\vert H\left( \bar{3}\right) ^{j}\right\vert 
\mathcal{B}_{i^{\prime }i^{\prime \prime }}\right\rangle \varepsilon
^{ii^{\prime }i^{\prime \prime }} \\
&&+T^{J}_{3}\left\langle \mathcal{P}_{k}^{i}\mathcal{M}_{j}^{k}\left\vert
H\left( \bar{3}\right) ^{j}\right\vert \mathcal{B}_{i^{\prime }i^{\prime
\prime }}\right\rangle \varepsilon ^{ii^{\prime }i^{\prime \prime
}}+T^{J}_{4}\left\langle \mathcal{P}_{j^{\prime }}^{i^{\prime }}\mathcal{M}%
_{j}^{i}\left\vert H\left( \bar{3}\right) ^{i^{\prime \prime }}\right\vert 
\mathcal{B}_{jj^{\prime }}\right\rangle \varepsilon _{ii^{\prime }i^{\prime
\prime }} \\
&&+T^{J}_{5}\left\langle \mathcal{P}_{j^{\prime }}^{i^{\prime }}\mathcal{M}%
_{j}^{i}\left\vert H\left( \bar{3}\right) ^{j}\right\vert \mathcal{B}%
_{i^{\prime \prime }j^{\prime }}\right\rangle \varepsilon _{ii^{\prime
}i^{\prime \prime }}+T^{J}_{6}\left\langle \mathcal{P}_{j}^{i^{\prime }}\mathcal{%
M}_{j^{\prime }}^{i}\left\vert H\left( \bar{3}\right) ^{j}\right\vert 
\mathcal{B}_{i^{\prime \prime }j^{\prime }}\right\rangle \varepsilon
_{ii^{\prime }i^{\prime \prime }},
\end{eqnarray*}%

\begin{figure*}[ht]
\begin{tabular}{cc}
\includegraphics[width=68mm,angle=0]{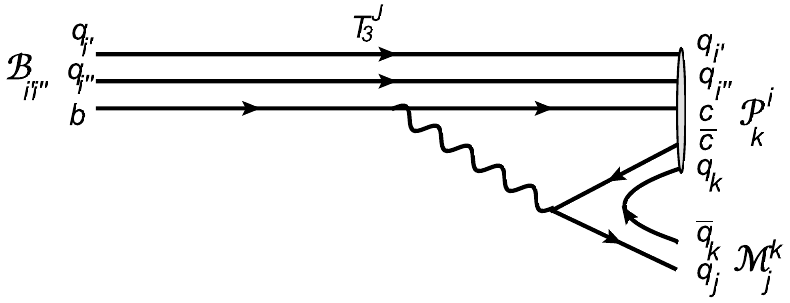} \ \ \
& \ \ \ \includegraphics[width=68mm,angle=0]{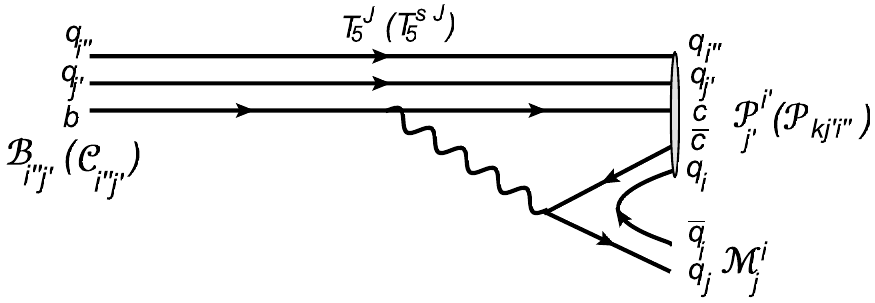}\\
\hspace{0.4cm}($\mathbf{a}$)&\hspace{0.8cm}($\mathbf{b}$)\\
& \\
\includegraphics[width=68mm,angle=0]{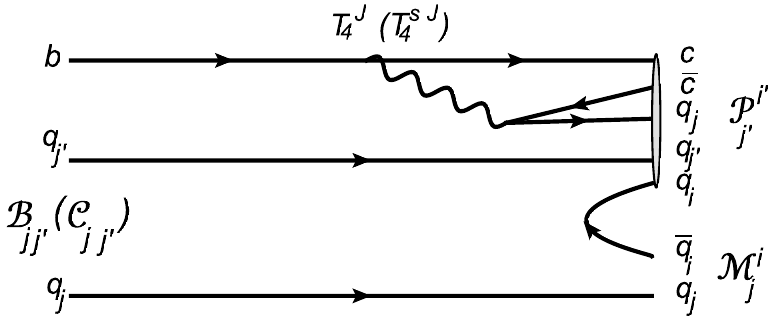} \ \ \
& \ \ \ \includegraphics[width=68mm,angle=0]{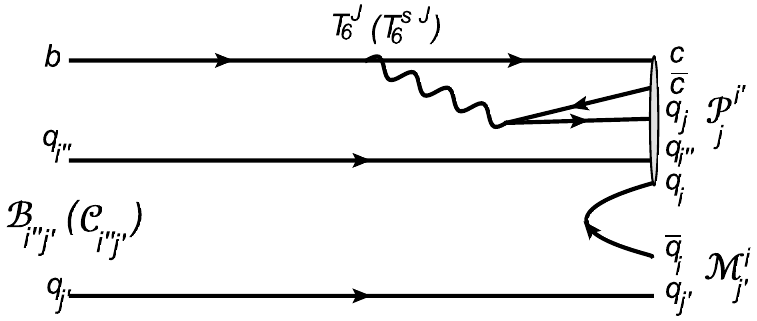}\\
\hspace{0.6cm}($\mathbf{c}$)&\hspace{1.2cm}($\mathbf{d}$)
\end{tabular}
\caption{ \sf Decays of  the  anti-triplet $(\mathcal{B}_{ij})$ and sextet $(\mathcal{C}_{ij})$ $b$-baryons into a 
hidden charm octet pentaquark state $\mathcal{P}_{i}^{j} $ and an octet of
pseudoscalar meson $\mathcal{M}_{i}^{j} $.
Figures (a) and (b) correspond to the case when the spin of initial state light diquark remains unchanged and is transferred to the pentaquark state. In Figures (c) and (d), the spin of  the initial state light diquark is shared between the  pentaquark and the meson in the final state  and their contribution is  suppressed in the heavy quark limit.} \label{production-1}
\end{figure*}

\noindent
where the superscript $J$ represents the spin of the final state pentaquark, $J=\frac{3}{2}$ or $J=\frac{5}{2}$.

For the  case of an anti-triplet $b$-baryon  decaying into a pentaquark state, the relevant contributions arise with the coefficients $T^{J}_1, T^{J}_2, T^{J}_3$ and $T^{J}_5$, in which the  spin of the initial state diquark in the decaying $b$-baryon remains the same in the final state pentaquark states. The contributions with the coefficients 
$T^{J}_4$ and $T^{J}_6$ arise only if the initial light diquark spin is shared by the pentaquark and the meson in the final state as shown in Figs. \ref{production-1}. These contributions are 
 suppressed by the heavy quark  symmetry.
 The coefficients $T^{J}_1$ and $T^{J}_2$ are expected to be smaller than $T^{J}_3$ and $T^{J}_5$,
as they involve higher Fock states of the pentaquarks (such as five quarks and two antiquarks).
The remaining dominant external $W$-emission diagrams  with the coefficients  $T^{J}_3$ and $T^{J}_5$ are shown in Figs. \ref{production-1} (a) and (b), respectively.  Noting that the amplitudes of the anti-triplet $b$-baryons decaying into an octet pentaquark and a pseudoscalar meson  enter in the combination $2T^{J}_3 + T^{J}_5$,  we denote this by $T^{J}_8$.  
So, for this class of decays there is just one amplitude for a given $J^P$ of the pentaquarks.
 Similarly, in the case of the sextet $b$-baryons decaying into decuplet pentaquark states, the only relevant diagram with external $W$-emission comes with $T_5^{s\; J}$. The
internal $W$-emission diagrams with the coefficients $T_4^{s\; J}$ and $T_6^{s\; J}$, shown in Fig. \ref{production-1} (c) and (d),  respectively, arise only if the initial light diquark spin is shared by the pentaquark and the meson in the final state. These contributions are again
 suppressed by the heavy quark  symmetry.   The amplitude for the sextet $b$-baryons decaying into a decuplet pentaquark and a meson is:
 \begin{equation}
\mathcal{A}^{J}_{t10}\left( q\right)  =T_{5}^{s\; J}\left\langle \mathcal{P}_{k j^{\prime}{i^{\prime\prime}}}\mathcal{M}%
_{l}^{k}\left\vert H\left( \bar{3}\right) ^{l}\right\vert \mathcal{C}%
_{i^{\prime\prime}j^{\prime }}\right\rangle \label{sextet-amplitude}.
\end{equation}%
 Also, in this class of decays, there is only one amplitude per $J^P$ due to heavy quark symmetry. An example is the decay of
 of the $\Omega_b^-=b (ss)$, which is an $SU(3)_F$ sextet with  the
$(ss)$ quarks having $j^p=1^+$. Due to heavy quark symmetry, $\Omega_b^-$ decay will produce a decuplet pentaquark, but not an
octet, though the SU(3) algebra admits the octet 
$3 \times 6= 8 + 10$. In particular, the amplitude  $\mathcal{A}^{J}_{s8}$ given in Eq. (13) in \cite{Li-He-He} is not allowed in the heavy quark symmetry limit. This underscores the use of the heavy quark symmetry  in $b$-baryonic decays to pentaquarks.

\begin{table*}[tbp]
\caption{\sf $SU(3)$ amplitudes corresponding to the $\Delta S = 0$ transitions for the $\Lambda _{b},\; \Xi _{b}^{0},\; \Xi _{b}^{-}$ and $ \Omega _{b}^{-}$ decays.  The symbols $\{X_{i} ; Y_{i}\}_{c_{i}}$  with $i = 1, ..., 5$  represent the different $\{\mathcal{P}_{X_i} ; \mathcal{P}_{Y_i}\}$ states with $c_{i}$ defining the flavor contents of the pentaquark states. For the decays of $\Lambda _{b},\; \Xi _{b}^{0},\; \Xi _{b}^{-}$ ($ \Omega _{b}^{-}$) the  pentaquark states are
 octet (decuplet).}
\label{DSzero-decays}
\begin{tabular}{|c|c|c|c|c|}
\hline
Decay Mode & Amplitude & Decay Mode & Amplitude \\ 
\hline
$\Lambda _{b}\rightarrow P_{p}^{\{X_{2}\; ; Y_{2}\}_{c_{1}}} \pi ^{-}$  & $%
T^{J}_{8}$ &  $\Xi _{b}^{-}\rightarrow P_{\Xi ^{-}}^{\{X_{2}\; ; Y_{2}\}_{c_{4}}} K^{0}$ & $%
T^{J}_{8}$  \\ 
\hline
$\Lambda _{b}\rightarrow P_{n}^{\{X_{2}\; ; Y_{2}\}_{c_{1}}} \eta _{8}$ &  $\frac{1}{%
\sqrt{6}}T^{J}_{8} $ & $\Xi _{b}^{-}\rightarrow P_{\Sigma ^{-}}^{\{X_{2}\; ; Y_{2}\}_{c_{2}}} \eta _{8}$ & 
$\frac{1}{\sqrt{6}}T^{J}_{8} $ \\ 
\hline
$\Lambda _{b}\rightarrow P_{n}^{\{X_{2}\; ; Y_{2}\}_{c_{1}}} \pi ^{0}$ &$-\frac{1}{%
\sqrt{2}}T^{J}_{8} $ & $\Xi _{b}^{-}\rightarrow P_{\Lambda ^{0}}^{\{X_{2}\; ; Y_{2}\}_{c_{2}}} \pi ^{-}$ & 
$\frac{1}{\sqrt{6}}T^{J}_{8} $ \\ 
\hline
$\Xi _{b}^{0}\rightarrow P_{\Lambda ^{0}}^{\{X_{2}\; ; Y_{2}\}_{c_{2}}} \eta _{8}$ & $-\frac{1}{6}T^{J}_{8}$  & $\Xi _{b}^{-}\rightarrow P_{\Sigma ^{0}}^{\{X_{2}\; ; Y_{2}\}_{c_{2}}}  \pi ^{-}$ & $%
\frac{1}{\sqrt{2}}T^{J}_{8} $ \\ 
\hline
$\Xi _{b}^{0}\rightarrow P_{\Sigma ^{0}}^{\{X_{2}\; ; Y_{2}\}_{c_{2}}}\eta _{8}$ & $\frac{1}{\sqrt{12}}T^{J}_{8} $ & $\Xi _{b}^{-}\rightarrow P_{\Sigma ^{-}}^{\{X_{2}\; ; Y_{2}\}_{c_{2}}} \pi ^{0}$ & $%
-\frac{1}{\sqrt{2}}T^{J}_{8} $  \\ \hline 
$\Xi _{b}^{0}\rightarrow P_{\Lambda ^{0}}^{\{X_{2}\; ; Y_{2}\}_{c_{2}}} \pi ^{0}$  & $\frac{1}{\sqrt{12}}T^{J}_{8}$ & $\Xi _{b}^{0}\rightarrow P_{\Sigma ^{0}}^{\{X_{2}\; ; Y_{2}\}_{c_{2}}} \pi ^{0}$ & $%
-\frac{1}{2}T^{J}_{8}$ \\
\hline
$\Omega _{b}^{-}\rightarrow P_{\Xi_{10} ^{-}}^{\{X_{3}\; ; Y_{3}\}_{c_{5}}} \pi^ {0}$ & 
$\frac{1}{\sqrt{6}}\left( -T_{5}^{s\; J}\right) $ & $\Omega _{b}^{-}\rightarrow P_{\Xi_{10}^{0}}^{\{X_{3}\; ; Y_{3}\}_{c_{5}}} \pi ^{-}$ & $%
\frac{1}{\sqrt{3}}T_{5}^{s\; J}$  \\
%$\Omega _{b}^{-}\rightarrow P_{\Omega^{-}_{10}}^{\{X_{3}\; ; Y_{3}\}_{c_{5}}} K^{0}$ & $T_{5}^{s}$ & &   \\ 
\hline \hline%
\end{tabular}%
\end{table*}

\begin{table*}[tbp]
\caption{\sf $SU(3)$ amplitudes corresponding to the $\Delta S = 1$  transitions for the $\Lambda _{b},\; \Xi _{b}^{0},\; \Xi _{b}^{-}$ and $ \Omega _{b}^{-}$ decays.  The symbols $\{X_{i} ; Y_{i}\}_{c_{i}}$
are the same as defined in Table \ref{DSzero-decays}.}
\label{DSone-decays}
\begin{tabular}{|c|c|c|c|c|}
\hline
Decay Mode & Amplitude & Decay Mode & Amplitude \\ 
\hline
$\Lambda _{b}\rightarrow P_{p}^{\{X_{2}\; ; Y_{2}\}_{c_{1}}}K^{-}$ & $%
T^{J}_{8} $  &  $\Xi _{b}^{-}\rightarrow P_{\Sigma ^{-}}^{\{X_{2}\; ; Y_{2}\}_{c_{2}}}  \bar{K}^{0}$
& $T^{J}_{8} $   \\ 
\hline
$\Lambda _{b}\rightarrow P_{n}^{\{X_{2}\; ; Y_{2}\}_{c_{1}}}\bar{K}^{0}$ & $%
T^{J}_{8} $ & $\Xi _{b}^{-}\rightarrow P_{\Xi ^{-}}^{\{X_{2}\; ; Y_{2}\}_{c_{2}}}  \eta _{8}$ & $%
-\frac{2}{\sqrt{6}}T^{J}_{8}  $ \\ 
\hline
$\Lambda _{b}\rightarrow P_{\Sigma ^{0}}^{\{X_{2}\; ; Y_{2}\}_{c_{3}}} \pi ^{0}$ & $%
0$ & $\Xi _{b}^{-}\rightarrow P_{\Lambda ^{0}}^{\{X_{2}\; ; Y_{2}\}_{c_{2}}}  K^{-}$ & $%
\frac{1}{\sqrt{6}}T^{J}_{8}  $  \\ 
\hline
$\Lambda _{b}\rightarrow P_{\Lambda ^{0}}^{\{X_{2}\; ; Y_{2}\}_{c_{3}}}\eta _{8}$
& $\frac{2}{3}T^{J}_{8} $  & $\Xi _{b}^{-}\rightarrow P_{\Sigma ^{0}}^{\{X_{2}\; ; Y_{2}\}_{c_{2}}}  K^{-}$ & $%
\frac{1}{\sqrt{2}}T^{J}_{8}  $\\ 
\hline
$\Xi _{b}^{0}\rightarrow P_{\Sigma ^{+}}^{\{X_{2}\; ; Y_{2}\}_{c_{2}}} K^{-}$ & $%
-T^{J}_{8} $  & $\Xi _{b}^{0}\rightarrow P_{\Xi ^{0}}^{\{X_{2}\; ; Y_{2}\}_{c_{2}}} \eta _{8}$ & $%
\frac{2}{\sqrt{6}}T^{J}_{8} $  \\
 \hline 
$\Omega _{b}^{-}\rightarrow P_{\Xi_{10} ^{0}}^{\{X_{3}\; ; Y_{3}\}_{c_{5}}} K^{-}$ & $%
\frac{1}{\sqrt{3}}T_{5}^{s\; J}$ & $\Omega _{b}^{-}\rightarrow P_{\Xi ^{-}_{10}}^{\{X_{3}\; ; Y_{3}\}_{c_{5}}} \bar{K}^{0}$
& $\frac{1}{\sqrt{3}}T_{5}^{s\; J}$  \\
%$\Omega _{b}^{-}\rightarrow P_{\Omega^{-}_{10}}^{\{X_{3}\; ; Y_{3}\}_{c_{5}}} \eta_{8}$ & $%
%-\frac{\sqrt{2}}{\sqrt{3}}T_{5}^{s}$ & $\Omega _{b}^{-}\rightarrow P_{\Omega^{-}}^{\{X_{3}\; ; Y_{3}\}_{c_{5}}} \eta_{1}$
%& $\frac{1}{\sqrt{3}}T_{5}^{s}$  \\
\hline \hline%
\end{tabular}%
\end{table*}

\section{$\boldsymbol{SU(3)}$ relations for the Pentaquark States}

Relations among  the  various decay amplitudes using $SU(3)_F$ symmetry have been derived in \cite{Li-He-He}. In terms of the internal and external $W$-boson classification, they are given in \cite{Cheng-Chua}. In particular,
in \cite{Li-He-He}, they are written including all possible topologies. However, heavy quark symmetry  suppresses
some contributions as we have argued in the previous section. 
 Combining $SU(3)_F$-symmetry
with the heavy quark symmetry , these relations involving pentaquarks with
a definite $J^P$ are given in  Tables \ref{DSzero-decays} and \ref{DSone-decays}. They are strikingly predictive
and we write them below, though they can all be read from these tables.

The $SU(3)$ relations involving different anti-triplet $b$-baryon decays into an octet pentaquark state and
a meson  from the Cabibbo-suppressed part of the weak Hamiltonian  $H^W(\Delta S=0)$  are:
\begin{eqnarray}
&&\mathcal{A}^{J}_{t8}(\Lambda _{b} \to P_{p}^{\{X_{2}\; ; Y_{2}\}_{c_{1}}} \pi ^{-})=-\sqrt{2}\mathcal{A}^{J}_{t8}(\Lambda _{b}\rightarrow P_{n}^{\{X_{2}\; ; Y_{2}\}_{c_{1}}} \pi ^{0})=\sqrt{6}\mathcal{A}^{J}_{t8}(\Lambda _{b}\rightarrow P_{n}^{\{X_{2}\; ; Y_{2}\}_{c_{1}}} \eta _{8})\;,\nonumber \\
&&\mathcal{A}^{J}_{t8}(\Xi _{b}^{0}\rightarrow P_{\Lambda ^{0}}^{\{X_{2}\; ; Y_{2}\}_{c_{2}}} \pi ^{0})=-\mathcal{A}^{J}_{t8}(\Xi _{b}^{0}\rightarrow P_{\Sigma ^{0}}^{\{X_{2}\; ; Y_{2}\}_{c_{2}}}\eta _{8})\;, \nonumber \\
&&\mathcal{A}^{J}_{t8}(\Xi _{b}^{-}\rightarrow P_{\Xi ^{-}}^{\{X_{2}\; ; Y_{2}\}_{c_{4}}} K^{0}) =\sqrt{6}\mathcal{A}^{J}_{t8}(\Xi _{b}^{-}\rightarrow P_{\Sigma ^{-}}^{\{X_{2}\; ; Y_{2}\}_{c_{2}}} \eta _{8})=\sqrt{6}\mathcal{A}^{J}_{t8}(\Xi _{b}^{-}\rightarrow P_{\Lambda ^{0}}^{\{X_{2}\; ; Y_{2}\}_{c_{2}}} \pi^{-})\;, \nonumber \\
 &&\mathcal{A}^{J}_{t8}(\Xi _{b}^{-}\rightarrow P_{\Xi ^{-}}^{\{X_{2}\; ; Y_{2}\}_{c_{4}}} K^{0})=\sqrt{2} \mathcal{A}^{J}_{t8}(\Xi _{b}^{-}\rightarrow P_{\Sigma ^{0}}^{\{X_{2}\; ; Y_{2}\}_{c_{2}}} \pi^{-})= -\sqrt{2}\mathcal{A}^{J}_{t8}(\Xi _{b}^{-}\rightarrow P_{\Sigma ^{-}}^{\{X_{2}\; ; Y_{2}\}_{c_{2}}} \pi^{0})\;,\label{r1}
 \end{eqnarray}
whereas for the $\Delta S=1$, the relations can be written as follows
 \begin{eqnarray}
&&\mathcal{A}^{J}_{t8}(\Lambda _{b}\rightarrow P_{p}^{\{X_{2}\; ; Y_{2}\}_{c_{1}}}K^{-})=\mathcal{A}^{J}_{t8}(\Lambda _{b}\rightarrow P_{n}^{\{X_{2}\; ; Y_{2}\}_{c_{1}}}\bar{K}^{0})\;, \nonumber\\
&& \mathcal{A}^{J}_{t8}(\Xi _{b}^{0}\rightarrow P_{\Sigma ^{+}}^{\{X_{2}\; ; Y_{2}\}_{c_{2}}} K^{-})=-\sqrt{\frac{3}{2}} \mathcal{A}^{J}_{t8}(\Xi _{b}^{0}\rightarrow P_{\Xi ^{0}}^{\{X_{2}\; ; Y_{2}\}_{c_{2}}} \eta _{8})= \mathcal{A}^{J}_{t8}(\Xi _{b}^{-}\rightarrow P_{\Sigma ^{-}}^{\{X_{2}\; ; Y_{2}\}_{c_{2}}}  \bar{K}^{0}),\nonumber \\
&&=\sqrt{2}\mathcal{A}^{J}_{t8}(\Xi _{b}^{-}\rightarrow P_{\Sigma ^{0}}^{\{X_{2}\; ; Y_{2}\}_{c_{2}}}  K^{-})\;.\label{r1s1}
\end{eqnarray}

Likewise, the amplitudes corresponding to the $\Delta S=0$ transitions of  the $b$-baryon sextet ($\Omega^{-}_b$) decaying into a decuplet pentaquark and a meson  are related:
\begin{equation}
 -\sqrt{2}\mathcal{A}^{J}_{t10}(\Omega _{b}^{-}\rightarrow P_{\Xi_{10} ^{-}}^{\{X_{3}\; ; Y_{3}\}_{c_{5}}} \pi^ {0})=\mathcal{A}^{J}_{t10}(\Omega _{b}^{-}\rightarrow P_{\Xi_{10} ^{0}}^{\{X_{3}\; ; Y_{3}\}_{c_{5}}} \pi^{-}), \label{r1}
\end{equation}
 and for $\Delta S = 1$, they corresponding ones are:
 \begin{equation}
\mathcal{A}^{J}_{t10}(\Omega _{b}^{-}\rightarrow P_{\Xi_{10} ^{0}}^{\{X_{3}\; ; Y_{3}\}_{c_{5}}} K^{-})= \mathcal{A}^{J}_{t10}(\Omega _{b}^{-}\rightarrow P_{\Xi_{10} ^{-}}^{\{X_{3}\; ; Y_{3}\}_{c_{5}}} \bar{K}^{0})\;.\label{r2} 
\end{equation}
Here, as before, $\mathcal{P}^{X_i}$ and $\mathcal{P}^{Y_i}$  have
 $J=\frac{3}{2}$ and $J=\frac{5}{2}$ , respectively. 

The two-body decay rate in the center-of-mass frame can be expressed as 
\begin{equation}
\Gamma \propto |q_{cm}| |\mathcal{A}|^2 \propto |q_{cm}|^{2L+1}, \label{rates}
\end{equation} 
where $|\mathcal{A}|$ is the amplitude of the respective decay mode,
 $L$ is the relative orbital angular momentum of the final state particles, and 
$q_{cm}$ is the center-of-mass momentum, defined as 
\begin{eqnarray}
|q_{cm}| &=& q_{\mathcal{P}} = \sqrt{E^2_{\mathcal{P}}-m^2_{\mathcal{P}}},\notag \\
E_{\mathcal{P}} &=& \frac{m^2_{\mathcal{B}}+m^2_{\mathcal{P}}-m^2_{\mathcal{M}}}{2m_{\mathcal{B}}}, 
 \label{cm-momentum}
\end{eqnarray}
where, $m_{\mathcal{B}}$,  $m_{\mathcal{P}}$ and $m_{\mathcal{M}}$ are the masses of the initial state $b$-baryon, final state pentaquark and pseudoscalar meson, respectively. Using this formula, the decay ratios $\Gamma(\mathcal{B}(\mathcal{C})\to \mathcal{P}^{5/2}\mathcal{M})/\Gamma(\Lambda^{0}_b \to P_{p}^{5/2}K^{-})$ for $\Delta S = 1$ and $\Delta S= 0$ transitions are summarized in the Tables \ref{Relative-Rates1} and \ref{Relative-Rates2}, respectively. 

In line with the arguments presented here,  we have argued that the state $P_c(4380)$ with $J^P=3/2^-$ is  an unlikely candidate for
a pentaquark. Rather, it is the state  with a lower mass, $P_c(4110)$ with $J^P=3/2^-$, which has the correct
angular momentum and light diquark spin to be produced in $\Lambda_b$ decays. The  ratios
    $\Gamma(\mathcal{B}(\mathcal{C})\to \mathcal{P}^{3/2}\mathcal{M})/\Gamma(\Lambda^{0}_b \to P_{p}^{{\{X_2\}}_{c_1}}K^{-})$ for $\Delta S = 1$ and $\Delta S= 0$ transitions are given in Tables \ref{Relative-Rates11} and \ref{Relative-Rates12}, respectively. In working out the numbers, we have used the pentaquark mass spectrum
    worked out by us. The small rates for the $\Delta S=0$ transitions compared to the  $\Delta S=1$
    reflect Cabibbo-suppression. In particular,  
    $\Gamma(\Lambda_b^0 \to {\mathcal P}_c(4450)^+ \pi^-)/\Gamma(\Lambda_b^0 \to {\mathcal P}_c(4450)^+ K^-)
    \simeq 0.08 $, and a similar number for the ratio involving the lower-mass $J^P=3/2^-$ state.\footnote{First evidence of 
 tetraquark- and pentaquark-structures  in the Cabibbo-suppressed channel 
 $\Lambda_b^0 \to J/\psi\; p\; \pi^-$ is reported
 recently by LHCb \cite{Aaij:2016ymb}.}

    In the numerical estimates
      for the $\Omega^{-}_b$ decays, we have assumed $T_{5}^{s\; J}= T^{J}_8$, which is expected as both of them are tree-amplitudes. However, this is only a ball-park estimate,  as the corresponding amplitudes, which are independent
      quantities, may differ substantially. 
     Many of the predictions presented here for the ratios can be sharpened by calculating the $SU(3)_F$-breaking and
     including the sub-leading contributions in $1/m_b$. More data on pentaquark states is required to estimate them.
      Absolute decay rates, on the other hand, require a reliable computation of 
     the amplitudes $T^{J}_8$ and $T_{5}^{s\; J} $. This, however,  is a daunting task, way beyond the
     theoretical tools available currently.

 \begin{table*}[tbp]
\caption{\sf {Estimate of the ratios of the decay widths  $\Gamma(\mathcal{B}(\mathcal{C})\to \mathcal{P}^{5/2}\mathcal{M})/\Gamma(\Lambda^{0}_b \to P_{p}^{5/2}K^{-})$ for $\Delta S = 1$ transitions \cite{Aaij:2015tga}.}}
\label{Relative-Rates1}
\begin{tabular}{|c|c|c|c|c|}
\hline
Decay Process & $\Gamma/\Gamma(\Lambda^{0}_b \to P_{p}^{5/2}K^{-})$ & Decay Process & $\Gamma/\Gamma(\Lambda^{0}_b \to P_{p}^{5/2}K^{-})$ \\ 
\hline
$\Lambda _{b}\rightarrow P_{p}^{\{Y_{2}\}_{c_{1}}}K^{-}$ & $%
1$  &  $\Xi _{b}^{-}\rightarrow P_{\Sigma ^{-}}^{\{Y_{2}\}_{c_{2}}}  \bar{K}^{0}$
& $2.07$   \\ 
\hline
$\Lambda _{b}\rightarrow P_{n}^{\{(Y_{2}\}_{c_{1}}}\bar{K}^{0}$ & $%
1$ & $\Xi _{b}^{0}\rightarrow P_{\Sigma ^{+}}^{\{Y_{2}\}_{c_{2}}} K^{-}$ & $%
2.07$ \\ 
\hline
$\Lambda _{b}\rightarrow P_{\Lambda^{0}}^{\{Y_{2}\}_{c_{3}}} \eta^{\prime}$ 
& $0.03 $  & $\Lambda _{b}\rightarrow P_{\Lambda^{0}}^{\{Y_{2}\}_{c_{3}}} \eta$ 
& $0.19 $ \\ 
\hline
 $\Xi _{b}^{-}\rightarrow P_{\Sigma^{0}}^{\{Y_{2}\}_{c_{2}}}  K^{-}$ & $%
1.04$ & $\Xi _{b}^{-}\rightarrow P_{\Lambda ^{0}}^{\{Y_{2}\}_{c_{2}}}  K^{-}$ & $%
0.34$  \\ 
\hline
$\Omega _{b}^{-}\rightarrow P_{\Xi_{10} ^{-}}^{\{Y_{3}\}_{c_{5}}} \bar{K}^{0}$
& $0.14$  & $\Omega _{b}^{-}\rightarrow P_{\Xi_{10} ^{0}}^{\{Y_{3}\}_{c_{5}}} K^{-}$ & $%
0.14$  \\
\hline \hline%
\end{tabular}%
\end{table*}
%

%%%%
\begin{table*}[tbp]
\caption{\sf {Estimate of the ratios of  decay widths $\Gamma(\mathcal{B}(\mathcal{C})\to \mathcal{P}^{5/2}\mathcal{M})/\Gamma(\Lambda^{0}_b \to P_{p}^{5/2}K^{-})$ for $\Delta S = 0$ transitions. These transitions are suppressed by a factor $|V^{\ast}_{cd}/V^{\ast}_{cs}|^2$ compared to $\Delta S = 1$ transitions \cite{Aaij:2015tga}.}}
\label{Relative-Rates2}
\begin{tabular}{|c|c|c|c|c|}
\hline
Decay Process & $\Gamma/\Gamma(\Lambda^{0}_b \to P_{p}^{5/2}K^{-})$ & Decay Process & $\Gamma/\Gamma(\Lambda^{0}_b \to P_{p}^{5/2}K^{-})$ \\ 
\hline
$\Lambda _{b}\rightarrow P_{p}^{\{Y_{2}\}_{c_{1}}} \pi ^{-}$  & $%
0.08$ &$\Lambda _{b}\rightarrow P_{n}^{\{Y_{2}\}_{c_{1}}} \pi ^{0}$ &$0.04  $   \\ 
\hline
$\Lambda _{b}\rightarrow P_{n}^{\{Y_{2}\}_{c_{1}}} \eta$  & $%
0.01$ &$\Lambda _{b}\rightarrow P_{n}^{\{Y_{2}\}_{c_{1}}} \eta^{\prime}$ &$0 $   \\ 
\hline
 $\Xi _{b}^{-}\rightarrow P_{\Xi ^{-}}^{\{Y_{2}\}_{c_{4}}} K^{0}$ & $0.02$ & $\Xi _{b}^{-}\rightarrow P_{\Sigma ^{0}}^{\{Y_{2}\}_{c_{2}}}  \pi ^{-}$ & $0.08$ \\ 
\hline
 $\Xi _{b}^{-}\rightarrow P_{\Sigma ^{-}}^{\{Y_{2}\}_{c_{2}}} \eta$ &$0.02$ &$\Xi _{b}^{-}\rightarrow P_{\Sigma ^{-}}^{\{Y_{2}\}_{c_{2}}} \eta^{\prime}$ &$0.01 $ \\ 
\hline
$\Xi _{b}^{-}\rightarrow P_{\Sigma ^{-}}^{\{Y_{2}\}_{c_{2}}} \pi ^{0}$ & $%
0.08$  & $\Xi _{b}^{0}\rightarrow P_{\Sigma ^{0}}^{\{Y_{2}\}_{c_{2}}} \pi ^{0}$ & $%
0.04$  \\ 
\hline
$\Xi _{b}^{0}\rightarrow P_{\Lambda ^{0}}^{\{X_{2}\ (Y_{2})\}_{c_{2}}} \eta$  & $0.01$ & $\Xi _{b}^{0}\rightarrow P_{\Lambda ^{0}}^{\{Y_{2}\}_{c_{2}}} \eta^{\prime}$ & $%
0.01$  \\
\hline
$\Xi _{b}^{0}\rightarrow P_{\Lambda ^{0}}^{\{Y_{2}\}_{c_{2}}} \pi ^{0}$  & $0.01$ & $\Omega _{b}^{-}\rightarrow P_{\Xi_{10} ^{-}}^{\{Y_{3}\}_{c_{5}}} \pi ^{0}$ & $%
0.01$  \\
\hline
$\Omega _{b}^{-}\rightarrow P_{\Xi_{10}^{0}}^{\{Y_{3}\}_{c_{5}}} \pi ^{-}$ & $%
0.02$ & &\\ \hline \hline%
\end{tabular}%
\end{table*}
%%%%%%%%%%%%%

 \begin{table*}[tbp]
\caption{\sf {Estimate of the ratios of decay widths $\Gamma(\mathcal{B}(\mathcal{C}) \to \mathcal{P}^{3/2}\mathcal{M})/\Gamma(\Lambda^{0}_b \to P_{p}^{{\{X_2\}}_{c_1}}K^{-})$ for $\Delta S = 1$ transitions. 
Note that we have used the pentaquark masses worked out in this work. }}
\label{Relative-Rates11}
\begin{tabular}{|c|c|c|c|c|}
\hline
Decay Process & $\Gamma/\Gamma(\Lambda^{0}_b \to P_{p}^{{\{X_2\}}_{c_1}K^{-}})$ & Decay Process & $\Gamma/\Gamma(\Lambda^{0}_b \to P_{p}^{{\{X_2\}}_{c_1}}K^{-})$ \\ 
\hline
$\Lambda _{b}\rightarrow P_{p}^{\{X_{2}\}_{c_{1}}}K^{-}$ & $%
1$  &  $\Xi _{b}^{-}\rightarrow P_{\Sigma ^{-}}^{\{X_{2}\}_{c_{2}}}  \bar{K}^{0}$
& $1.38 $   \\ 
\hline
$\Lambda _{b}\rightarrow P_{n}^{\{X_{2}\}_{c_{1}}}\bar{K}^{0}$ & $%
1 $ & $\Xi _{b}^{0}\rightarrow P_{\Sigma ^{+}}^{\{X_{2}\}_{c_{2}}} K^{-}$ & $%
1.38$ \\ 
\hline
$\Lambda _{b}\rightarrow P_{\Lambda^{0}}^{\{X_{2}\}_{c_{3}}} \eta^{\prime}$ 
& $0.17 $  & $\Lambda _{b}\rightarrow P_{\Lambda^{0}}^{\{X_{2}\}_{c_{3}}} \eta$ 
& $0.22 $ \\ 
\hline
 $\Xi _{b}^{-}\rightarrow P_{\Sigma^{0}}^{\{X_{2}\}_{c_{2}}}  K^{-}$ & $%
0.69$ & $\Xi _{b}^{-}\rightarrow P_{\Lambda ^{0}}^{\{X_{2}\}_{c_{2}}}  K^{-}$ & $%
0.23$  \\ 
\hline
$\Omega _{b}^{-}\rightarrow P_{\Xi_{10} ^{-}}^{\{X_{3}\}_{c_{5}}} \bar{K}^{0}$
& $0.24$  & $\Omega _{b}^{-}\rightarrow P_{\Xi_{10} ^{0}}^{\{X_{3}\}_{c_{5}}} K^{-}$ & $%
0.24$  \\
\hline \hline%
\end{tabular}%
\end{table*}
%

%%%%
\begin{table*}[tbp]
\caption{\sf {Estimate of the ratios of the decay widths  $\Gamma(\mathcal{B}(\mathcal{C}) \to \mathcal{P}^{3/2}\mathcal{M})/\Gamma(\Lambda^{0}_b \to P_{p}^{{\{X_2\}}_{c_1}}K^{-})$ for $\Delta S = 0$ transitions. These transitions are suppressed by a factor $|V^{\ast}_{cd}/V^{\ast}_{cs}|^2$ compared to $\Delta S = 1$ transitions. 
Other input values are the same as in Table \ref{Relative-Rates11}. }}
\label{Relative-Rates12}
\begin{tabular}{|c|c|c|c|c|}
\hline
Decay Process & $\Gamma/\Gamma(\Lambda^{0}_b \to P_{p}^{{\{X_2\}}_{c_1}K^{-}})$ & Decay Process & $\Gamma/\Gamma(\Lambda^{0}_b \to P_{p}^{{\{X_2\}}_{c_1}}K^{-})$ \\ 
\hline
$\Lambda _{b}\rightarrow P_{p}^{\{X_{2}\}_{c_{1}}} \pi ^{-}$  & $%
0.06$ &$\Lambda _{b}\rightarrow P_{n}^{\{X_{2}\}_{c_{1}}} \pi ^{0}$ &$0.03  $   \\ 
\hline
$\Lambda _{b}\rightarrow P_{n}^{\{X_{2}\}_{c_{1}}} \eta$  & $%
0.01 $ &$\Lambda _{b}\rightarrow P_{n}^{\{X_{2}\}_{c_{1}}} \eta^{\prime}$ &$0.01 $   \\ 
\hline
 $\Xi _{b}^{-}\rightarrow P_{\Xi ^{-}}^{\{X_{2}\}_{c_{4}}} K^{0}$ & $0.02$ & $\Xi _{b}^{-}\rightarrow P_{\Sigma ^{0}}^{\{X_{2}\}_{c_{2}}}  \pi ^{-}$ & $0.03$ \\ 
\hline
 $\Xi _{b}^{-}\rightarrow P_{\Sigma ^{-}}^{\{X_{2}\}_{c_{2}}} \eta$ &$0.02$ &$\Xi _{b}^{-}\rightarrow P_{\Sigma ^{-}}^{\{X_{2}\}_{c_{2}}} \eta^{\prime}$ &$0.01 $ \\ 
\hline
$\Xi _{b}^{-}\rightarrow P_{\Sigma ^{-}}^{\{X_{2}\}_{c_{2}}} \pi ^{0}$ & $%
0.04$  & $\Xi _{b}^{0}\rightarrow P_{\Sigma ^{0}}^{\{X_{2}\}_{c_{2}}} \pi ^{0}$ & $%
0.02$  \\ 
\hline
$\Xi _{b}^{0}\rightarrow P_{\Lambda ^{0}}^{\{X_{2}\}_{c_{2}}} \eta$  & $0$ & $\Xi _{b}^{0}\rightarrow P_{\Lambda ^{0}}^{\{X_{2}\}_{c_{2}}} \eta^{\prime}$ & $%
0$  \\
\hline
$\Xi _{b}^{0}\rightarrow P_{\Lambda ^{0}}^{\{X_{2}\}_{c_{2}}} \pi ^{0}$  & $0.01$ & $\Omega _{b}^{-}\rightarrow P_{\Xi_{10} ^{-}}^{\{X_{3}\}_{c_{5}}} \pi ^{0}$ & $%
0.01$  \\
\hline
$\Omega _{b}^{-}\rightarrow P_{\Xi_{10}^{0}}^{\{X_{3}\}_{c_{5}}} \pi ^{-}$ & $%
0.02 $ & &\\ \hline \hline%
\end{tabular}%
\end{table*}
%%%%%%%%%%%%%%%%%%%%%%%%%%%%%%%%%%%%%%%%%%%%%%%%%%%%%%%%%%%%%%%%%%%%%%%%%%%%%%%%%%%%%%%%%%%%%%%%%%%%%%%%%%%%%%%%%%%%%%%%%%%%%%%%%%%%%%%%
\section{Concluding Remarks}
We have studied the mass spectrum of the pentaquarks $\bar{c}[cq][q^\prime
q^{\prime\prime}]$ where $q, q^{\prime}, q^{\prime \prime}$ can be
$u$,  $d$, and $s$ quarks, in the diquark-diquark-antiquark approach, using an effective Hamiltonian.
The various input parameters are fixed from the experimentally measured states $X, Y,  Z$,
under the assumption that they are diquark-antidiquark states. The determination of the spin-spin couplings in
a diquark differs between the  type I~\cite{Maiani:2004vq} and type II~\cite{Maiani:2014aja} models,  and we have worked out the numerics in both cases.
 Our mass estimates for the pentaquark states agree with the ones given in \cite{Maiani:2015vwa}, using
 their spin and angular-momentum assignments: $P_c^+(4380)= \{\bar{c} [cu]_{s=1} [ud]_{s=1}; L_\mathcal{P}=0,
J^{\rm P}=\frac{3}{2}^- \}$ and  $P_c^+(4450)= \{\bar{c} [cu]_{s=1} [ud]_{s=0}; L_\mathcal{P}=1,
J^{\rm P}=\frac{5}{2}^+ \}$. They correspond to the $S$-wave  pentaquark state
 $\mathcal{P}_{X_4}$ and the $P$-wave pentaquark state $\mathcal{P}_{Y_2}$, respectively, in our
 work and given in Table \ref{Table I}. We have argued that as the state $\mathcal{P}_{X_4}$ has a light diquark
 with spin 1, $[ud]_{s=1}$,  heavy-quark symmetry suppresses the decay 
 $\Lambda_b^0 \to \mathcal{P}_{X_4} K^-$. Hence, identifying the lower-mass state $P_c^+(4380)$ with the state 
 $\mathcal{P}_{X_4}$, as done in \cite{Maiani:2015vwa} is, in our opinion,  problematic.
 In other words, only those $b$-baryon decays in which the spin of the light diquark is transferred to the spin of
 the light diquark in the pentaquark go unhindered. 
 
  However, there is a $J^P=\frac{3}{2}^-$ state in the spectrum, called 
 $\mathcal{P}_{X_2}= \{\bar{c} [cu]_{s=1} [ud]_{s=0}; L_\mathcal{P}=0, J^{\rm P}=\frac{3}{2}^- \}$  in Table \ref{Table I}, which has the right diquark-spin to be produced in $\Lambda_b$ decays.
 The mass difference between the  $J^P=\frac{5}{2}^+$ pentaquark state $P_c(4450)$ and the $J^P=\frac{3}{2}^-$
 pentaquark state $\mathcal{P}_{X_2}$ can be determined by the corresponding mass difference in
 the charm-baryon sector $m[\Lambda_c^+(2625); J^P=\frac{3}{2}^-] - m[\Lambda_c^+(2286); J^P=\frac{1}{2}^+] \simeq 341$ 
 MeV.  Using the effective Hamiltonian, and fixing the parameters from the tetraquark states, we estimate this
 mass difference to be about 320 MeV,  thus yielding the mass of the $J^P=\frac{3}{2}^-$ pentaquark in the range
 4110 - 4130 MeV. We urge the LHCb collaboration to reanalyze their data to search for this lower-mass
 $ J^P=\frac{3}{2}^- $  pentaquark state decaying into $J/\psi\; p$. 
 
 The pentaquark spectrum  with hidden $c\bar{c}$ and three light quarks is very rich. Restricting to the
 lowest $S$- and $P$-waves, there are 50 such states in the mass range 4100 - 5100 MeV. They can,
 in principle, be produced in prompt processes at the LHC but face unfavorable rates and
 formidable background.  A small fraction of these
 pentaquarks having the correct flavor and spin/angular-momentum quantum numbers can be produced in $b$-baryonic decays and their searches appear very promising.   We have studied  these decays in the second part of our paper, following closely
 earlier analysis along the same lines \cite{Cheng-Chua, Li-He-He}. However, there is one difference
 between our work and theirs in that we use the heavy quark symmetry to classify the various topologies (diagrams),
 and have kept only those transitions which are allowed by
 the heavy-quark-spin conservation. This substantially reduces  the number of allowed amplitudes
 and results in a number of relations which can be tested in the future.
  Relative  rates
 for certain decay channels are calculated and we find that there are good chances to discover yet other
 pentaquark states in the decays of the $b$ baryons  $\Xi _{b}^{0}\ , \Xi _{b}^{-}$ and $\Omega_{b}$.
 In the meanwhile, we await for more data and renewed analysis of the current LHCb data to test some of the predictions presented here,
 in particular, the existence of a $J^P=3/2^-$ pentaquark state having a mass around 4110 MeV 
 in the Dalitz analysis of the decay $\Lambda_b \to J/\psi p K^-$.
 
\textbf{Acknowledgments}

One of us (A.A.) would like to thank Luciano Maiani, Antonello Polosa and Sheldon Stone for helpful discussions.
I.~Ahmed, J.~Aslam and A.~Rehman acknowledge discussions with  Prof. Fayyazuddin.

\end{document}